\documentclass{article}
\usepackage{amsfonts}
\usepackage{proof}
\usepackage{color}
\usepackage{graphicx}
\newcommand*{\fnot}{\sim\!}
\newcommand*{\limp}{\Rightarrow}
\newcommand*{\imp}{\supset}
\newtheorem{definition}{Definition}
\newtheorem{theorem}[definition]{Theorem}
\newtheorem{lemma}[definition]{Lemma}
\newtheorem{example}[definition]{Example}
\newtheorem{proposition}[definition]{Proposition}
\newtheorem{corollary}[definition]{Corollary}

\title{\textbf{Sequent and Hypersequent Calculi for Abelian and \L ukasiewicz Logics}}
\author{George Metcalfe$^1$, Nicola Olivetti$^2$ and Dov Gabbay$^1$\\
\small{$^1$Department of Computer Science, King's College London, Strand, London}\\
\small{WC2R 2LS, UK email:\{metcalfe,dg\}@dcs.kcl.ac.uk}\\
\small{$^2$Department of Computer Science, University of Turin, Corso Svizzera 185,}\\
\small{10149 Turin, Italy email:olivetti@di.unito.it}}
\date{}
\begin{document}
\maketitle
\begin{abstract}
\noindent
We present two embeddings of infinite-valued \L ukasiewicz logic \textbf{\L} into Meyer and Slaney's abelian logic \textbf{A}, the logic of lattice-ordered abelian groups. We give new analytic proof systems for \textbf{A} and use the embeddings to derive corresponding systems for \textbf{\L}. These include: hypersequent calculi for \textbf{A} and \textbf{\L} and terminating versions of these calculi; labelled single sequent calculi for \textbf{A} and \textbf{\L} of complexity co-NP; unlabelled single sequent calculi for \textbf{A} and 
\textbf{\L}.
\end{abstract}
\newcommand{\hide}[1]{#1}
                 
\section{Introduction}
In \cite{haj:met} H\'ajek classifies \emph{truth-functional fuzzy logics} as logics whose conjunction and implication functions are interpreted via \emph{continuous t-norms} and their residua. Since each continuous t-norm may be constructed from the \L ukasiewicz, G\"odel and Product t-norms, the resulting logics, \L ukasiewicz logic \textbf{\L}, G\"odel logic \textbf{G} and Product logic \textbf{P} respectively, are fundamental for this classification. Also important is H\'ajek's  Basic Logic \textbf{BL} \cite{haj:met} which was proved in \cite{CEGT99} to characterise validity in logics based on continuous t-norms. In a similar vein Godo and Esteva \cite{est:mtl} have defined the logics \textbf{MTL} (Monoidal T-norm based Logic) and \textbf{IMTL} (Involutive Monoidal T-norm based Logic)  which turn out to characterise validity in \emph{left-continuous t-norm based logics} and \emph{left-continuous t-norm with involutive negation based logics} respectively. Underlying all the above systems is H\"ohle's monoidal logic \textbf{ML} \cite{hoh:com} which provides a common basis for both t-norm logics and logics based on Heyting algebras. A diagram of the relationships between these logics is given in Figure 1 with arrows signifying inclusions between logics.

\begin{figure}[htbp]
\input{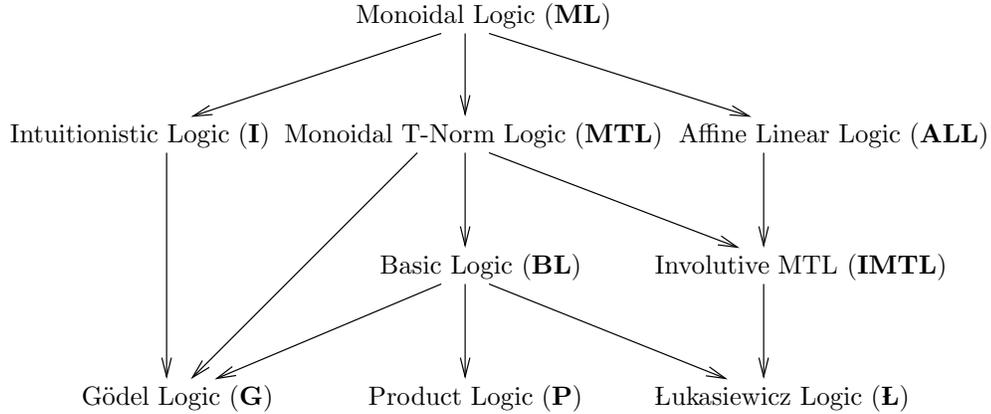}
\caption{Relationships between fuzzy logics}
\label{fig1}
\end{figure}
Systematic accounts of Hilbert-style axiomatisations and algebraic semantics for fuzzy logics have been presented eg in \cite{Gottwald99,haj:met} but no comprehensive Gentzen-style sequent calculus approach  such as that given for substructural logics  has been forthcoming. Progress for each logic has varied. \L ukasiewicz logic \textbf{\L}, despite having been explored thoroughly from a semantic perspective (see eg \cite{cig:alg}), has been given only calculi that either fail to be analytic (ie cut-free) \cite{Prijatelj96,Ciabattoni:1997:TCB} or internal (ie avoiding non-logical calculations) \cite{hahnle:admvl,mun:res,oli:tab}. A notable exception is the work of Aguzzoli, Ciabattoni and Gerla \cite{AguzzoliCiabattoni99,AguzzoliGerla02} who give cut-free internal calculi for \textbf{\L}, \textbf{G} and \textbf{P}; their approach however is based on validity in \emph{finite-valued} logics with sequents consisting of components representing sets of truth values, and therefore does not integrate well with more standard Gentzen presentations. For G\"odel logic \textbf{G} the proof-theoretic picture is more healthy; there are several (rather complicated) single sequent calculi \cite{son:gen,AFM:dupftc,Dyckhoff99,AvrKon:00}, a calculus employing \emph{sequents of relations} \cite{Baaz:1999:ACP}, and also a very natural \emph{hypersequent} formulation 
\cite{Avron96a}. Hypersequent calculi, introduced independently by Avron in \cite{Avron87} and Pottinger in \cite{Pottinger83} as a generalisation of the usual sequent calculi, have also been given for \textbf{MTL} and other variants of Urquhart's \textbf{C} logic \cite{urqu:many86} in \cite{baa:pro,Ciabattoni00a}. A first semantics-based calculus for H\'ajek's \emph{Basic Logic} \textbf{BL} has been presented in \cite{mon:tab}.

The approach taken in this work towards developing a proof theory of fuzzy logics is new. It results from the identification of fuzzy logics with \emph{fragments of comparative logics}. Intuitively, a fuzzy logic with truth values between $0$ and $1$ may be viewed as being part of an ``extended'' logic 
with truth values between $-\infty$ and $+\infty$. These extended logics are called comparative logics\footnotemark \ here to distinguish them from fuzzy logics with truth values in $[0,1]$. The idea proposed in this paper is that (at least in some cases) comparative logics are more natural to work with for proof-theoretic purposes. Developing proof systems for comparative logics then allows us to exploit our translations to obtain proof systems for 
their fragments, the fuzzy logics. In fact this mirrors the strategy of systems with extra relation symbols or labels; calculations are performed in an extended language to find answers for queries in the original language. The novelty here is that this approach is followed at the \emph{logical} level rather than the \emph{metalogical} level.

\footnotetext{The expression \emph{comparative} logic was first used by Casari 
in \cite{cas:com,cas:ab} to characterise a set of systems formalising ideas expressed by Aristotle about comparisons of majority, minority 
and equality. There is considerable overlap between Casari's comparative logics and those presented here but whereas Casari regards fuzzy logics such as \L ukasiewicz infinite-valued logic as particular comparative 
logics, this work instead presents fuzzy logics as \emph{fragments} of 
comparative logics.}

We apply this approach here to \L ukasiewicz infinite-valued logic \textbf{\L}. For \textbf{\L} the appropriate comparative logic is \emph{abelian} logic \textbf{A}, the logic of lattice-ordered abelian groups with characteristic models in the integers, rationals and reals. \textbf{A} was introduced and motivated independently by Meyer and Slaney \cite{mey:ab} and Casari \cite{cas:com} as a relevance and comparative logic respectively. In \cite{mey:ab} the so-called \emph{enthymematic} fragment of \textbf{A} was proved to coincide exactly with \textbf{\L}$\mathbf{^+}$, the positive part of \textbf{\L} and this result is extended here to identify the \emph{material} fragment of \textbf{A} with the whole of \textbf{\L}. A further, perhaps more natural, translation from \textbf{\L} into \textbf{A} is also provided. By introducing analytic sequent and hypersequent calculi for \textbf{A} (the first proof systems for this logic) we are then able to derive the first Gentzen-style analytic sequent and hypersequent calculi for \textbf{\L}.

We proceed as follows. In section 2 we introduce \L ukasiewicz logic \textbf{\L} and Abelian logic \textbf{A} presenting some important and relevant results. In section 3 we relate the 
logics via two embeddings of \textbf{\L} into \textbf{A}. We then turn our attention to proof theory. In section 4 we present \emph{hypersequent calculi} for \textbf{A} and \textbf{\L} and in section 5 we introduce \emph{terminating} versions of these calculi. In section 6 we introduce \emph{labelled calculi} for \textbf{A} and \textbf{\L} which we show to be co-NP. Finally in section 7 we present \emph{unlabelled single sequent calculi} for \textbf{A} and \textbf{\L}.\footnotemark
\footnotetext{Some of the results presented in sections 6 and 7 have already appeared in \cite{met:ana}.}
\section{Background}
In this section we first present some algebraic preliminaries and then introduce \L ukasiewicz infinite-valued logic \textbf{\L} and abelian logic \textbf{A}.
\subsection{Preliminaries}
The work reported here is concerned with \emph{propositional} logics; the 
language used is built inductively from a denumerable set of propositional 
variables $p_1, p_2, \ldots$ and a finite set of connectives.
\begin{definition}[$\mathbf{Con}$]
$Con=\{\land,\lor,\oplus,+,\imp,\to,\limp,\leftrightarrow,\lnot,\fnot,t,\bot\}$
\end{definition}
\begin{definition}[$\mathbf{For}$]
$For$ is built inductively as follows: (1) $p_1,p_2,\ldots \in For$, (2) if $\phi_1, \ldots, \phi_m \in For$ then $*(\phi_1, 
\ldots, \phi_m) \in For$ where $* \in Con$, $arity(*)=m$.
\end{definition}
\begin{definition}[A-valuation]
Given an algebra $A = \langle L,O \rangle$ with carrier $L \neq \emptyset$ and operations $\{t\} \subseteq O \subseteq Con$, an 
A-valuation is a function $v:For \rightarrow L$ such that for $* \in Con$, 
$arity(*)=m$, $v(*(\phi_1,\ldots,\phi_m)) = *(v(\phi_1),\ldots,v(\phi_m))$.
\end{definition}
\begin{definition}[Validity]
$\phi \in For$ is valid in an algebra $A$ iff $v(\phi) \ge t$ for all 
$A$-valuations $v$. $\phi$ is X-valid 
(written $\models_X \phi$) iff $\phi$ is valid in all algebras of the class $X$.
\end{definition}
\subsection{\L ukasiewicz Logic}
\L ukasiewicz infinite-valued logic \textbf{\L} was introduced by \L ukasiewicz in 1930 \cite{LukUnt}. Good references for historical details and an overview of the main results are \cite{Malinowski93} and \cite{urqu:many86}. An in-depth algebraic treatment of \textbf{\L} is provided by \cite{cig:alg} and the fuzzy logic perspective is described in \cite{haj:met}.
\begin{definition}[\L ukasiewicz Infinite-Valued Logic, \L] $\supset$ and $\bot$ 
are primitive. \textbf{\L} has the rule $(mp)$ and the following definitions and 
axioms:
\[\begin{array}[t]{llll}
D\!\fnot & \fnot A = A \supset \bot
& D\!\lor & A \lor B = (A \supset B) \supset B\\
Dt & t = \fnot \bot 
& D\!\oplus & A \oplus B = \fnot A \supset B\\
\L 1 & A \supset (B \supset A)
& \L 3 & ((A \supset B) \supset B) \supset ((B \supset A) \supset A)\\
\L 2 & (A \supset B) \supset ((B \supset C) \supset (A \supset C))
& \L 4 & ((A \supset \bot) \supset (B \supset \bot)) \supset (B \supset A)
\end{array}\]
\[
\newcommand*{\modp}{\begin{array}[t]{c}
A \supset B, A\\
\hline
B
\end{array}}
\begin{array}[t]{rl}
(mp) & \modp
\end{array}
\]
\end{definition}
The following axiomatisation of \emph{positive} \L ukasiewicz logic was given by Rose and Rosser in \cite{roserosser:58}.
\begin{definition}[Positive \L ukasiewicz Infinite-Valued Logic, 
\textbf{\L}$\mathbf{^+}$]
$\supseteq$, $\lor$, $\land$ and $t$ are primitive. \textbf{\L}$\mathbf{^+}$ has 
the rule (mp) and the following axioms:
\[\begin{array}[t]{llll}
\L^+1 & (D \supseteq B) \supseteq ((D \supseteq C) \supseteq (D \supseteq (B 
\land C)))
& \L^+6 & (B \supseteq C) \lor (C \supseteq B)\\
\L^+2 & B \supseteq (C \supseteq B)
& \L^+7 & (B \land C) \supseteq B\\
\L^+3 & (B \supseteq C) \supseteq ((C \supseteq D) \supseteq (B \supseteq D))
& \L^+8 & (B \land C) \supseteq C\\
\L^+4 & ((B \supseteq C) \supseteq C) \supseteq (B \lor C)
& \L^+9 & t\\
\L^+5 & (B \lor C) \supseteq (C \lor B) & &
\end{array}\]
\end{definition}
Algebraic structures for \textbf{\L} were introduced by Chang in 1958 \cite{Chang58}.
\begin{definition}[MV-algebra]
An MV-algebra is an algebra $\langle A,\oplus,\fnot,\bot \rangle$ with a binary 
operation $\oplus$, a unary operation $\fnot$ and a constant $\bot$, 
satisfying the following equations:
\[\begin{array}[t]{llll}
mv1 & \bot \oplus  a = a & 
mv2 & a \oplus b = b \oplus a\\
mv3 & (a \oplus b) \oplus c = a \oplus (b \oplus c) & 
mv4 & \fnot \fnot a = a\\
mv5 & a \oplus \fnot \bot = \fnot \bot &
mv6 & \fnot(\fnot a \oplus b) \oplus b = \fnot(\fnot b \oplus a) \oplus a
\end{array}\]
We also define: $a \supset b = \fnot a \oplus b$, $t = \fnot \bot$, $a \lor b = \fnot (\fnot a \land \fnot b)$.
\end{definition}
Let $[0,1]_\mathbb{R}$ be the real unit interval and define $a \oplus b = 
min(1,a+b)$, $\fnot a = 1-a$ and $\bot=0$. We have that $[0,1]_{\textrm{\L}} = \langle [0,1], \oplus, \fnot, \bot \rangle$ is an MV-algebra with $a \supset b = min(1,1+b-a)$. In fact $[0,1]_\textrm{\L}$ is \emph{characteristic} for MV-algebras.
\begin{theorem}[Chang 1958 \cite{Chang58}] 
The following are equivalent:
(1) $\phi$ is a theorem of \textbf{\L}.
(2) $\phi$ is valid in all MV-algebras.
(3) $\phi$ is valid in $[0,1]_{\textrm{\L}}$.
\end{theorem}
It will be convenient for us to consider the (also characteristic) 
MV-algebra $[-1,0]_{\textrm{\L}} = \langle [-1,0]_\mathbb{R}, \oplus, \fnot, \bot \rangle$ where $a \oplus b = min(0,a+b+1)$, $\fnot a = -1 - a$ and $\bot = -1$ with $a \supset b = min(0,b-a)$.

Finally we mention the following complexity result for \textbf{\L}.
\begin{theorem}[Mundici 1987 \cite{Mundici:1987:SMS}]
The tautology problem for \textbf{\L} is co-NP-complete.
\end{theorem}
\subsection{Abelian Logic}
Abelian logic \textbf{A} was introduced and investigated by Meyer and Slaney 
\cite{mey:ab,mey:ab2} as a logic of \emph{relevance}, obtained from Anderson and Belnap's relevant logic \textbf{R} (see \cite{AB75} for details) by  
rejecting the contraction axiom $((A \to (A \to B)) \to (A \to B)$ and 
liberalising the negation axiom $((A \to \bot) \to \bot) \to A$ to $((A \to B) 
\to B) \to A$. \textbf{A} has also been motivated independently by Casari \cite{cas:ab} as a logic of \emph{comparison} formalising comparisons of majority, minority and equality in natural language. More recently Galli et al. \cite{gallewsag:log} have derived \textbf{A} as a logic of 
equilibrium for arguments. Finally we note that a sequent calculus for the \emph{intensional} fragment of \textbf{A} (the logic of abelian groups) has been provided by Paoli in \cite{pao:log}.

We begin here by giving Meyer and Slaney's axiomatisation of \textbf{A}.
\begin{definition}[Abelian Logic, A] $\to, +, \land, \lor$ and $t$ are 
primitive. \textbf{A} 
has the following definitions, axioms and rules:
\[\begin{array}[t]{llll}
D\!\leftrightarrow & A \leftrightarrow B = (A \to B) \land (B \to A)
& D\!\lnot & \lnot A = A \to t\\
A1 & ((A \lor B) \to C) \leftrightarrow ((A \to C) \land (B \to C))
& A6 & A \leftrightarrow (t \to A)\\
A2 & ((A+B) \to C) \leftrightarrow (A \to (B \to C))
& A7 & (A \land B) \to A\\
A3 & (A \to B) \to ((B \to C) \to (A \to C))
& A8 & (A \land B) \to B\\
A4 & ((A \to B) \land (A \to C)) \to (A \to (B \land C))
& A9 & A \to ((A \to B) \to B)\\
A5 & (A \land (B \lor C)) \to ((A \land B) \lor (A \land C))
& A10 & ((A \to B) \to B) \to A
\end{array}\]
\[
\newcommand*{\modp}{\begin{array}[t]{c}
A \to B, A\\
\hline
B
\end{array}}
\newcommand*{\Iand}{\begin{array}[t]{c}
A, B\\
\hline
A \land B
\end{array}}
\begin{array}[t]{cccc}
(mp) & \modp & \ \ (\land I) & \Iand
\end{array}\]
\end{definition}
The appropriate class of algebras for \textbf{A} are \emph{lattice-ordered abelian groups}:
\begin{definition}[Lattice-Ordered Abelian Group (Abelian L-Group)]
An abelian l-group is an algebra $\langle G,+,\lor,\lnot,t \rangle$ with 
binary operations $+$ and $\lor$, a unary operation $\lnot$ and a constant 
$t$, satisfying the following equations:
\[\begin{array}[t]{llll}
a1 & t+a = a & a5 & a \lor b = b \lor a\\
a2 & a+b = b+a & a6 & (a \lor b) \lor c = a \lor (b \lor c)\\
a3 & (a+b)+c = a+(b+c) & a7 & a=a \lor a\\
a4 & a +\lnot a=t & a8 & a+(b \lor c) = (a+b) \lor (a+c)
\end{array}\]
In addition, we define: $a \land b = \lnot (\lnot a \lor \lnot b)$, $a \to b = 
\lnot a + b$, $a \leftrightarrow b = (a \to b) \land (b \to a)$ and $a \le b$ iff $a \lor b = b$.
\end{definition}
Well known examples of abelian l-groups are the integers 
$\mathbb{Z} = \langle 
\mathbb{Z},+,max,-,0 \rangle$, the rationals $\mathbb{Q} = \langle 
\mathbb{Q},+,max,-,0 \rangle$ and the 
reals $\mathbb{R} = \langle \mathbb{R},+,max,-,0 \rangle$. In fact any of 
these serves as a characteristic 
model for \textbf{A}.
\begin{theorem}[Characterisation Theorem for A]
The following are equivalent: 
(1) $\phi$ is a theorem of \textbf{A}. 
(2) $\phi$ is valid in all abelian l-groups. 
(3) $\phi$ is valid in $\mathbb{Z}$.
(4) $\phi$ is valid in $\mathbb{Q}$.
(5) $\phi$ is valid in $\mathbb{R}$.
\end{theorem}
\textbf{Proof.} The equivalence of (2)-(5) is well known from pure mathematics. 
The equivalence of (1) and (2) 
is proved by Meyer and Slaney in \cite{mey:ab}; also given there is a more logic 
minded proof of the equivalence of 
(2) and (3). $\Box$\\[.1in]
Notice that in \textbf{A} we have $a + b = \lnot(\lnot a + \lnot b)$ ie 
intensional \emph{disjunction} and intensional 
\emph{conjunction} are exactly the same thing. Similarly we have that $t = \lnot t$ so truth 
and canonical falsity are also 
identical.

The complexity of the tautology problem for \textbf{A} follows directly from the following theorem.
\begin{theorem}[Weispfenning \cite{Weispfenning:1986:CWP}]
The word problem for abelian l-groups is co-NP-complete.
\end{theorem}
\begin{corollary}
The tautology problem for \textbf{A} is co-NP-complete.
\end{corollary}
\section{Relating A and \L}
The use of abelian l-groups in the theory of \L ukasiewicz infinite-valued logic is not new. Chang's original completeness proof for \textbf{\L} involved extending MV-algebras to abelian l-groups and more recently Cignoli et al. have proved that MV-algebras and abelian l-groups are \emph{categorically equivalent} \cite{cig:alg}. Ordered abelian groups are also fundamental to the Kripke-style semantics for \textbf{\L} provided independently by Scott \cite{scott:74} and Urquhart \cite{urqu:many86}. It is not surprising therefore that the \emph{logic} of abelian l-groups and \textbf{\L} should be related in some way. Here we give both an embedding of \textbf{\L} into \textbf{A} which mirrors relationships between other logics investigated by Meyer and co-workers, and also a translation from \textbf{\L} into \textbf{A} which will prove useful later in deriving calculi for \textbf{\L}.

We begin by defining two new implications for \textbf{A}.
\begin{definition}[Enthymematic Implication ($\supseteq$)]
$A \supseteq B = (t \land A) \to B$
\end{definition}
\begin{definition}[Material Implication ($\supset$)]
$A \supset B = (t \land A) \to (\bot \lor B)$
\end{definition}
In \cite{mey:int} Meyer shows that fragments obtained using these new implications frequently correspond to other well-known logics; for example the material and enthymematic fragments of \textbf{R} are classical logic and  
positive intuitionistic logic respectively. Of particular interest here is the 
fact, demonstrated by Dunn and Meyer in \cite{dunn:alge71}, that the material fragment of G\"odel logic \textbf{G} is the relevance logic \textbf{RM}. We now investigate the corresponding fragments for \textbf{A}, noting that since \textbf{A} has no falsity constant, $\bot$ is treated as a \emph{propositional variable}.
\begin{definition}[Enthymematic Fragment of A, $\mathbf{A_E}$] $\mathbf{A_E}$ 
consists of the theorems of $\mathbf{A}$ in the language built up from propositional variables, $t$, $\land$, $\lor$ and $\supseteq$.
\end{definition}
\begin{definition}[Material Fragment of A, $\mathbf{A_M}$] $\mathbf{A_M}$ 
consists of the theorems of $\mathbf{A}$ in the language built up from propositional variables, $\bot$ and $\supset$.
\end{definition}
Meyer and Slaney in \cite{mey:ab} show that the enythmematic fragment of 
\textbf{A} is \textbf{\L}$\mathbf{^+}$, the positive part of \L ukasiewicz logic. We generalise this here to identify the material fragment of \textbf{A} with the whole of \textbf{\L}.
\begin{theorem}[Meyer and Slaney \cite{mey:ab}]
\textbf{\L}$\mathbf{^+}=\mathbf{A_E}$
\end{theorem}
\begin{theorem}
\textbf{\L}$ \ = \mathbf{A_M}$
\end{theorem}
\textbf{Proof.}\\[.1in]
\textbf{\L}$ \ \subseteq \mathbf{A_M}$. It is straightforward if laborious to 
check that all the (translated) axioms of 
\textbf{\L} are valid in $\mathbb{Q}$ so it remains to show that $(mp)$ is 
admissible. Consider $\mathbf{A_M}$ 
formulae, $\phi$ and $\psi$ where $\models_A \phi \supset \psi$ and $\models_A 
\phi$. Since $\models_A \phi$ we have 
$\phi \ge t$, whence $\phi \supset \psi = (\phi \land t) \to (\psi \lor \bot) = 
\psi \lor \bot \ge t$. If $\psi$ is a 
propositional variable then by taking a valuation $v(\psi)=v(\bot)=-1$ in 
$\mathbb{Q}$ we have $v(\phi \supset 
\psi) < t$, a contradiction. So either $\psi=t$ or $\psi=\psi_1 \supset \psi_2$. 
If the former then clearly 
$\models_A \psi$, if the latter then we have $\psi = (\psi_1 \land t) \to (\psi_2 \lor \bot) \ge \psi_2 \lor \bot \ge \bot$ whence $\psi \lor \bot 
= \psi 
\ge t$ and $\models_A \psi$ as 
required.\\[.1in]
$\mathbf{A_M} \subseteq$ \textbf{\L}.  We show that if $\not 
\models_{\textrm{\L}} \phi$ then $\not \models_A \phi$. Given a valuation $v$ for $[-1,0]_\textbf{\L}$ such that $v(\phi) < 0$, define a 
valuation $v'$ in $\mathbb{Q}$ as follows: $v'(\bot)=-1$, $v'(p) = 
v(p)$ 
for all propositional variables 
$p$.
We claim that for all formulae $\psi$: (1) if $v(\psi) < 0$ then $v'(\psi) = 
v(\psi)$, (2) if $v(\psi) = 0$ then $v'(\psi) \ge 0$. 
We would then have $v'(\phi) = v(\phi) < 0$ as required.
We prove the claim by induction on the complexity of $\psi$.
The base case holds by stipulation.
If $\psi = \psi_1 \supset \psi_2 = (t \land \psi_1) \to (\bot \lor \psi_2)$, 
then if $v(\psi_1) > v(\psi_2)$ we have 
$v(\psi) = v(\psi_2) - v(\psi_1)$. $v(\psi_2) < 0$ so $v'(\bot \lor 
\psi_2) 
= v(\psi_2)$. Also $v'(t \land 
\psi_1) = v(\psi_1)$, 
whence $v'(\psi) = v(\psi_2) - v(\psi_1)$. If $v(\psi_1) \le v(\psi_2)$ then 
if $v(\psi_2) = 0$ we have $v'(\bot \lor \psi_2) \ge 0$ by the induction 
hypothesis and $v'(t \land \psi_1) \le 0$ by definition, whence we have 
$v'(\psi) 
\ge 0$; if $v(\psi_2) < 0$ then $v(\psi_1) < 0$ and we have $v'(\psi_1) = 
v(\psi_1)$ and $v'(\psi_2) = v(\psi_2)$ whence $v'(\psi) = v(\psi_2) - 
v(\psi_1) \le 0$ as required. $\Box$\\[.1in]
We now give a less general but possibly more intuitive translation of \textbf{\L} into \textbf{A}. The idea is to restrict the valuations of atoms in \textbf{A} to lie between $t$ and an arbitary fixed propositional variable $q^\bot$ that acts as falsity. \L ukasiewicz implication (replacing $\supset$ by $\limp$ from now on) is defined in \textbf{A} as follows:
\begin{definition}[Positive Implication $\limp$]
$A \limp B = (A \to B) \land t$
\end{definition}
Before continuing we note that $\land$ and $\lor$ can be interpreted using only $\limp$ and the intensional connectives:
\begin{proposition} \label{proplimpdef}
(i) $a \land b = a + (a \limp b)$ (ii) $a \lor b = (b \limp a) \to a$
\end{proposition}
\textbf{Proof.} (i) $a + (a \limp b) = a + (t \land (a \to b)) = (a + t) \land 
(a+ (a \to b)) = 
a \land b$ (ii) $(b \limp a) \to a = \lnot((t \land (b \to a)) + \lnot a) = 
\lnot((t+\lnot a) 
\land ((b \to a) + \lnot a)) = \lnot(\lnot a \land (\lnot (b + \lnot a) + \lnot 
a)) = 
\lnot(\lnot a \land (\lnot b + (a + \lnot a))) = \lnot(\lnot a \land \lnot b) = a \lor b$. $\Box$\\[.1in]
Our translation of \textbf{\L} into \textbf{A} is given below:
\begin{definition}[Translation *] \label{deftrans}
$p^* = (p \lor q^\bot) \land t$, $\bot^* = q^\bot \land t$, $(A \limp B)^* = t \land (A^* \to B^*)$ where $q^\bot$ is an arbitary fixed propositional variable of \textbf{A}.
\end{definition}
\begin{theorem} \label{thmtrans}
$\models_\textrm{\L} \phi$ iff $\models_A \phi^*$.
\end{theorem}
\textbf{Proof.}\\[.1in]
For the left-to-right direction suppose that $v(\phi^*) < 0$ in $\mathbb{Q}$. If $v(q_\bot) \ge 0$ then it is easy to prove that $v(\phi^*)=0$ which is a contradiction. Hence we have that $v(q_\bot) < 0$ and in fact we can assume WLOG that $v(q_\bot) = -1$ (multiplying our valuation $v$ if necessary). We now define a valuation $v'$ for $[-1,0]_\textrm{\L}$ where $v'(p) = v(p^*)$. We prove inductively that for all \textbf{\L}-formulae $\psi$, $v'(\psi) = v(\psi^*)$ which gives us $v'(\phi) = v(\phi^*) < 0$ in $[-1,0]_\textrm{\L}$ as required. The base cases for propositional variables and $\bot$ follow by definition, for $\limp$ we have that $v'(\psi_1 \limp \psi_2) = min(0,v'(\psi_2)-v'(\psi_1)) = min(0,v(\psi_2^*) - v(\psi_1^*)) = v(t \land (\psi_1^* \to \psi_2^*)) = v((\psi_1 \limp \psi_2)^*)$ and we are done.\\[.1in]
For the right-to-left direction suppose that $v(\phi) < 0$ in $[-1,0]_\textrm{\L}$, we define a valuation $v'$ in $\mathbb{Q}$ such that $v'(p) = v(p)$ and $v'(q^\bot) = -1$ and prove inductively that $v'(\psi^*) = v(\psi)$ for 
all \textbf{\L}-formulae $\psi$, whence $v'(\phi) < 0$ in $\mathbb{Q}$ as required. The base cases follow immediately by definition and for $\limp$ we have $v'((\psi_1 \limp \psi_2)^*) = v'(t \land (\psi_1^* \to \psi_2^*)) = min(0,v'(\psi_2^*) - v'(\psi_1^*)) = min(0,v(\psi_2) - v(\psi_1)) = v(\psi_1 \limp \psi_2)$ and we are done. $\Box$
\section{Hypersequent Calculi}
Hypersequent calculi were introduced independently by Avron \cite{Avron87} and 
Pottinger \cite{Pottinger83} for logics with a reasonably simple semantics 
lacking a correspondingly simple sequent calculus. Hypersequents consist of multiple components (sequents) interpreted disjunctively and hypersequent rules include, in addition to single sequent rules, external structural rules that can operate on more than one component at a time. Hypersequent calculi are particularly suitable for logics obeying \emph{prelinearity} (ie $(A \to B) \lor (B \to A)$) since the multiple components of the hypersequent allow several alternative hypotheses to be processed in parallel. All t-norm based logics have prelinearity and this suggests that the hypersequent formalism could be a more suitable framework for a systematic proof theory of fuzzy logics than the usual Gentzen single-sided sequents; indeed natural and intuitive hypersequent calculi have already been provided for G\"odel logic \textbf{G} \cite{Avron91b} and \textbf{MTL} \cite{baa:pro,Ciabattoni00a}.
\begin{definition}[Hypersequent] 
A component is an ordered pair of multisets of formulae written $\Gamma \vdash \Delta$ ie a 
sequent; a hypersequent is a multiset of components written $\Gamma_1 \vdash 
\Delta_1|\ldots|\Gamma_n \vdash \Delta_n$.
\end{definition}
Note that we choose components to consist of pairs of \emph{multisets} of formulae and hypersequents to consist of \emph{multisets} of components and therefore avoid the need in our calculi for internal and external exchange rules. In what follows \emph{all} set terminology $\cup$, $\cap$, $\{\}$ and so on will refer to multisets. We will also make use of the following definitions.
\begin{definition}[Formula Complexity cp(A)]
$cp(q) = 0$ for $q$ atomic, \\$cp(*(A_1,\ldots,A_m)) = 1 + \Sigma_{i=1}^m cp(A_i)$ where $* \in Con$, $arity(*)=m$.
\end{definition}
\begin{definition}[Hypersequent Multiset Complexity mc(G)]
$mc(\Gamma_1 \vdash \Delta_1|\ldots|\Gamma_n \vdash \Delta_n) = \{cp(A) : A \in \cup_{i=1}^n \Gamma_i \cup \Delta_i\}$
\end{definition}
\begin{definition}[Integer Multiset Ordering $<_m$] For $\Gamma$ and $\Delta$ multisets of integers, $\Gamma <_m \Delta$ if either (1) $\Gamma \subset \Delta$ or (2) $\Gamma <_m \Delta'$ where $\Delta' = \Delta - \{j\} \cup \{i,\ldots,i\}$ and $i < j$.
\end{definition}
Note that $<_m$ is a \emph{well-order} on multisets of integers.
\subsection{A Hypersequent Calculus for A}
Hypersequents for \textbf{A} are interpreted in the standard way, recalling that intensional conjunction and intensional disjunction are equivalent in this logic:
\begin{definition}[Interpretation of Hypersequents for A]
Let the interpretation of a sequent $S=A_1,\ldots,A_n \vdash B_1,\ldots,B_m$ be $\phi^S=(A_1 + \ldots + A_n) \to (B_1 + \ldots + B_m)$ (where $C_1 + \ldots + C_k = t$ if $k=0$) so that $\models S$ iff $\models \phi^S$. Let the interpretation of a hypersequent $G=S_1|\ldots|S_n$ be $\phi^G=\phi^{S_1} \lor \ldots \lor \phi^{S_n}$ so that $\models G$ iff $\models \phi^G$.
\end{definition}
We introduce the following hypersequent calculus for \textbf{A}.
\begin{definition}[GA]
\textbf{GA} has the following rules:\\[.1in]
\small{
Axioms
\[
\begin{array}[t]{ccccc}
(ID) & A \vdash A & \ \ \ \ \ \ & (\Lambda) & \vdash
\end{array}\]
Structural rules
\[
\newcommand*{\extweak}{\begin{array}[t]{c}
G|\Gamma \vdash \Delta\\
\hline
G|\Gamma \vdash \Delta|\Gamma' \vdash \Delta'
\end{array}}
\newcommand*{\extcont}{\begin{array}[t]{c}
G|\Gamma \vdash \Delta|\Gamma \vdash \Delta\\
\hline
G|\Gamma \vdash \Delta
\end{array}}
\newcommand*{\extsc}{\begin{array}[t]{c}
G|\Gamma_1,\Gamma_2 \vdash \Delta_1,\Delta_2\\
\hline
G|\Gamma_1 \vdash \Delta_2|\Gamma_2 \vdash \Delta_2
\end{array}}
\newcommand*{\extmingle}{\begin{array}[t]{c}
G|\Gamma_1 \vdash \Delta_1 \ \ G|\Gamma_2 \vdash \Delta_2\\
\hline
G|\Gamma_1,\Gamma_2 \vdash \Delta_1,\Delta_2
\end{array}}
\begin{array}[t]{lclc}
(EW) & \extweak & (EC) & \extcont\\
& & &\\
(S) & \extsc & (M) & \extmingle
\end{array}
\]
Logical rules
\[
\newcommand*{\trightrule}{\begin{array}[t]{c}
G|\Gamma \vdash \Delta\\
\hline
G|\Gamma \vdash t, \Delta
\end{array}}
\newcommand*{\tleftrule}{\begin{array}[t]{c}
G|\Gamma \vdash \Delta\\
\hline
G|\Gamma,t \vdash \Delta
\end{array}}
\newcommand*{\arrowrightrule}{\begin{array}[t]{c}
G|\Gamma,A \vdash B,\Delta\\
\hline
G|\Gamma \vdash A \to B, \Delta
\end{array}}
\newcommand*{\arrowleftrule}{\begin{array}[t]{c}
G|\Gamma,B \vdash A,\Delta\\
\hline
G|\Gamma, A \to B \vdash \Delta
\end{array}}
\newcommand*{\plusrightrule}{\begin{array}[t]{c}
G|\Gamma \vdash A,B,\Delta\\
\hline
G|\Gamma \vdash A+B, \Delta
\end{array}}
\newcommand*{\plusleftrule}{\begin{array}[t]{c}
G|\Gamma,A,B \vdash \Delta\\
\hline
G|\Gamma, A+ B \vdash \Delta
\end{array}}
\newcommand*{\andrightrule}{\begin{array}[t]{c}
G|\Gamma \vdash A,\Delta \ \ G|\Gamma \vdash B,\Delta\\
\hline
G|\Gamma \vdash A \land B, \Delta
\end{array}}
\newcommand*{\andleftrule}{\begin{array}[t]{c}
G|\Gamma,A \vdash \Delta\\
\hline
G|\Gamma, A \land B \vdash \Delta
\end{array}}
\newcommand*{\orleftrule}{\begin{array}[t]{c}
G|\Gamma, A \vdash \Delta \ \ G|\Gamma, B \vdash \Delta\\
\hline
G|\Gamma, A \lor B \vdash \Delta
\end{array}}
\newcommand*{\orrightrule}{\begin{array}[t]{c}
G|\Gamma \vdash A, \Delta\\
\hline
G|\Gamma \vdash A \lor B \vdash \Delta
\end{array}}
\newcommand*{\neworrightrule}{\begin{array}[t]{c}
G|\Gamma \vdash A, \Delta|\Gamma \vdash B,\Delta\\
\hline
G|\Gamma \vdash A \lor B, \Delta
\end{array}}
\newcommand*{\newandleftrule}{\begin{array}[t]{c}
G|\Gamma,A \vdash \Delta|\Gamma,B \vdash \Delta\\
\hline
G|\Gamma, A \land B \vdash \Delta
\end{array}}
\newcommand*{\lognotleftrule}{\begin{array}[t]{c}
G|\Gamma \vdash A, \Delta\\
\hline
G|\Gamma, \lnot A \vdash \Delta
\end{array}}
\newcommand*{\lognotrightrule}{\begin{array}[t]{c}
G|\Gamma, A \vdash \Delta\\
\hline
G|\Gamma \vdash \lnot A, \Delta
\end{array}}
\begin{array}[t]{lclc}
(t,l) & \tleftrule & (t,r) & \trightrule\\
& & &\\
(\lnot,l) & \lognotleftrule & (\lnot,r) & \lognotrightrule\\
& & &\\
(\to,l) & \arrowleftrule & (\to,r) & \arrowrightrule\\
& & &\\
(+,l) & \plusleftrule & (+,r) & \plusrightrule\\
& & &\\
(\land,l) & \newandleftrule & (\land,r) & \andrightrule\\
& & &\\
(\lor,l) & \orleftrule & (\lor,r) & \neworrightrule
\end{array}
\]}
\end{definition}
Alternative (non-invertible) rules to replace $(\land,l)$ and $(\lor,r)$ without losing completeness are:
\[
\small{
\newcommand*{\andlefrule}{\begin{array}[t]{cl}
G|\Gamma,A_i \vdash \Delta\\
\hline
G|\Gamma, A_1 \land A_2 \vdash \Delta
\end{array}}
\newcommand*{\andleftrule}{\begin{array}[t]{ll}
\andlefrule & i=1,2
\end{array}}
\newcommand*{\orrighrule}{\begin{array}[t]{c}
G|\Gamma \vdash A_i, \Delta\\
\hline
G|\Gamma \vdash A_1 \lor A_2 \vdash \Delta
\end{array}}
\newcommand*{\orrightrule}{\begin{array}[t]{ll}
\orrighrule & i=1,2
\end{array}}
\begin{array}[t]{lclc}
(\land'_i,l) & \andleftrule & (\lor_i',r) & \orrightrule
\end{array}}
\]
We could also replace the rules $(M)$, $(EW)$ and the axioms $(ID)$, 
$(\Lambda)$, $(t,l)$, $(t,r)$ with the single axiom:
\[\small{
\begin{array}[t]{lc}
(AX) & G|\Gamma,t,\ldots,t \vdash \Gamma,t,\ldots,t
\end{array}}\]
Prelinearity for \textbf{A} is proved as follows.
\begin{example} \label{exama}
A proof of $(A \to B) \lor (B \to A)$ in \textbf{GA}:\\
\[\infer{\vdash (A \to B) \lor (B \to A)}{
	\infer{\vdash A \to B|\vdash B \to A}{
		\infer{A \vdash B|\vdash B \to A}{
			\infer{A \vdash B|B \vdash A}{
				\infer{A,B \vdash A,B}{
					A \vdash A & B \vdash B}}}}}
\]
\end{example}
We now turn our attention to the soundness and completeness of \textbf{GA}. Note that we use the convention in what follows of writing $\Gamma$ for both the multiset $\Gamma$ and the sum of elements of $\Gamma$. We also write $\lambda \Gamma$ for $\underbrace{\Gamma \cup \ldots \cup \Gamma}_\lambda$.
\begin{theorem}[Soundness of GA] \label{thmsoundga}
If $G$ succeeds in \textbf{GA} then $\models_\mathbf{A} G$.
\end{theorem}
\textbf{Proof.} We reason inductively on the height of a proof in \textbf{GA} showing that the axioms are sound and that the rules of \textbf{GA} preserve validity in $\mathbb{Q}$:
\begin{itemize}

\item 	Axioms. Clearly $A - A \ge 0$ and $0 \ge 0$

\item	Structural Rules. For $(S)$, if $\models_\mathbf{A} 
	G|\Gamma_1,\Gamma_2 \vdash \Delta_1,\Delta_2$ then either $G \ge 0$ 
	or $(\Delta_1 + \Delta_2) - (\Gamma_1 + \Gamma_2) \ge 0$. So we have 
	either $G \ge 0$ or $\Delta_1 - \Gamma_1 \ge 0$ or $\Delta_2 - \Gamma_2 
	\ge 0$ whence we have $\models_\mathbf{A} 
	G|\Gamma_1 \vdash \Delta_1|\Gamma_2 \vdash \Delta_2$. The other 
	structural rules are similar.

\item 	Logical Rules. For $(\to,r)$, if $\models_\mathbf{A} G|\Gamma,A 
	\vdash \Delta, B$ then either $G \ge 0$ or 
	$(\Delta + B) - (\Gamma + A) \ge 0$ 
	so either $G \ge 0$ or 
	$(\Delta + B - A) - \Gamma \ge 0$ 
	ie $\models_\mathbf{A} G|\Gamma \vdash \Delta, A \to B$. The other 
	logical rules are similar. $\Box$

\end{itemize}
We prepare for our completeness proof by proving some simple properties of \textbf{GA}.
\begin{proposition} \label{propid}
$\Pi \vdash \Pi$ succeeds in $\mathbf{GA}$
\end{proposition}
\textbf{Proof.} By induction on the size of $\Pi$. If $\Pi = \emptyset$ then we apply $(\Lambda)$. Otherwise $\Pi = 
\{A\} \cup \Pi'$ and we apply $(M)$ to obtain $A \vdash A$ which succeeds by $(ID)$, 
and $\Pi' \vdash \Pi'$ which 
succeeds by the induction hypothesis. $\Box$
\begin{proposition} \label{propterm}
The logical rules of \textbf{GA} are terminating.
\end{proposition}
\textbf{Proof.} It is easy to see that all the logical rules strictly reduce the multiset 
complexity of the hypersequent. $\Box$
\begin{definition}[Invertible] 
A rule is invertible if whenever the conclusion of the rule is valid then all 
its premises are valid.
\end{definition}
\begin{proposition} \label{propinv}
The logical rules of \textbf{GA} are invertible.
\end{proposition}
\textbf{Proof.} We reason in $\mathbb{Q}$. As an example consider $(\land,l)$; if 
$\models_\mathbf{A} G|\Gamma,A \vdash \Delta|\Gamma,B \vdash \Delta$ then 
either $G \ge 0$, $\Delta - (\Gamma + B) \ge 0$ or $\Delta - (\Gamma + A) \ge 
0$. So if $G < 0$ then we have 
$\Delta - (\Gamma + min(A,B)) \ge 0$ ie $\models_\mathbf{A} G|\Gamma,A \land B 
\vdash \Delta$. The other logical rules are 
similar. $\Box$\\[.1in]
We now show that proving an atomic hypersequent in \textbf{GA} is equivalent to solving a linear programming problem over $\mathbb{Q}$.
\begin{proposition} \label{proplin}
Given $\Gamma_i, \Delta_i$ containing only atoms for $i = 1 \ldots n$ then 
$\models_A \Gamma_1 \vdash 
\Delta_1|\ldots|\Gamma_n \vdash \Delta_n$ iff there exist 
$\lambda_1,\ldots,\lambda_n \in \mathbb{Z}^+$ such 
that $\lambda_i > 0$ for some $i$, $1 \le i \le n$ and $\cup_{i=1}^n \lambda_i 
\Gamma_i = \cup_{i=1}^n \lambda_i \Delta_i$.
\end{proposition}
\textbf{Proof.} Reasoning in $\mathbb{Q}$ we have $\models_A \Gamma_1 \vdash 
\Delta_1|\ldots|\Gamma_n \vdash \Delta_n$ iff $\Gamma_1 \le 
\Delta_1$ or $\ldots$ or $\Gamma_n \le \Delta_n$. This is 
equivalent to the set $\{\Gamma_1 > 
\Delta_1, \ldots, \Gamma_n > \Delta_n\}$ being inconsistent over 
$\mathbb{Q}$ which holds iff there exist 
$\lambda_1,\ldots,\lambda_n \in \mathbb{Z}^+$ such that $\lambda_i > 0$ for some 
$i$, $1 \le i \le n$ and $\Sigma_{i=1}^n \lambda_i \Gamma_i = \Sigma_{i=1}^n 
\lambda_i \Delta_i$ ie iff $\cup_{i=1}^n \lambda_i \Gamma_i = \cup_{i=1}^n 
\lambda_i \Delta_i$. $\Box$
\begin{proposition} \label{propcom}
If there exist $\lambda_1,\ldots,\lambda_n \in \mathbb{Z}^+$ such that 
$\cup_{i=1}^n \lambda_i \Gamma_i = 
\cup_{i=1}^n \lambda_i \Delta_i$ then $\Gamma_1 \vdash \Delta_1|\ldots|\Gamma_n 
\vdash \Delta_n$ succeeds in 
$\mathbf{GA}$.
\end{proposition}
\textbf{Proof.}
For each $i$, if $\lambda_i=0$ then we apply $(EW)$ to remove $\Gamma_i \vdash 
\Delta_i$ from the hypersequent, otherwise we apply 
$(EC)$ $\lambda_i - 1$ times to obtain $\lambda_i$ copies of $\Gamma_i \vdash 
\Delta_i$. We then apply $(S)$ 
repeatedly to get a component of the form $\Pi \vdash \Pi$ (since 
$\cup_{i=1}^n \lambda_i \Gamma_i = 
\cup_{i=1}^n \lambda_i \Delta_i$) which succeeds by Proposition \ref{propid}. $\Box$\\[.1in]
We are now ready to prove the completeness of \textbf{GA}.
\begin{theorem}[Completeness of GA] \label{thmcompga}
If $\models_\mathbf{A} G$ then $G$ succeeds in \textbf{GA}
\end{theorem}
\textbf{Proof.} We show that given a hypersequent $G$ valid in \textbf{A} we are 
able to find a proof of $G$ in \textbf{GA}. Our first step in the proof is to 
apply the logical rules exhaustively to $G$; since by Propositions \ref{propterm} and 
\ref{propinv} these rules are terminating and invertible respectively we obtain atomic hypersequents valid in \textbf{A}. It therefore suffices to show that each 
\emph{atomic} hypersequent $G' = \Gamma_1 
\vdash \Delta_1|\ldots|\Gamma_n \vdash \Delta_n$ valid in \textbf{A} is provable 
in \textbf{GA}. By Proposition \ref{proplin} we have that $G'$ is valid iff 
there exist 
$\lambda_1,\ldots,\lambda_n \in \mathbb{Z}^+$ such 
that $\lambda_i > 0$ for some $i$, $1 \le i \le n$ and $\cup_{i=1}^n \lambda_i 
\Gamma_i = \cup_{i=1}^n \lambda_i \Delta_i$ ie a linear combination of the 
components of $G'$ (taking $\lambda_i$ copies of each $\Gamma_i \vdash \Delta_i$ 
and adding them all together) gives a component of the form $\Pi \vdash \Pi$. So 
now by Proposition \ref{propcom}, which says that we can apply $(EC)$, $(EW)$ 
and $(S)$ to $G'$ to obtain the required combination, we have that $G'$ 
succeeds. $\Box$\\[.1in]
We note that Theorem \ref{thmcompga} tells us that the following \emph{cut rules} are admissible in \textbf{A}:
\[
\small{
\begin{array}[t]{ccc}
\infer{\Gamma \vdash \Delta}{
	\Gamma,A \vdash \Delta,A} & \ \ \ &
\infer{\Gamma, \Pi \vdash \Delta, \Sigma}{
	\Gamma,A \vdash \Delta & \Pi \vdash \Sigma,A}
\end{array}}
\]
In fact these are interderivable in \textbf{GA}:
\[
\small{
\begin{array}[t]{ccc}
\infer{\Gamma, \Pi \vdash \Delta, \Sigma}{
	\infer{\Gamma, \Pi, A \vdash \Delta, \Sigma, A}{
		\Gamma, A \vdash \Delta &
		\Pi \vdash \Sigma, A}} & \ \ \ &
\infer{\Gamma \vdash \Delta}{
	\infer{\Gamma \vdash \Delta, A \to A}{
		\Gamma, A \vdash \Delta, A} &
	\infer{A \to A \vdash}{
		A \vdash A}}
\end{array}}
\]
\subsection{A Hypersequent Calculus for \L}
We derive a hypersequent calculus for \textbf{\L} from \textbf{GA}, our hypersequent calculus for \textbf{A}, in several stages. First we give an interpretation for hypersequents for \textbf{\L} in terms of the characteristic model $[-1,0]_\textrm{\L}$. This allows us to give a very natural translation from hypersequents for \textbf{\L} to hypersequents for \textbf{A}. We then immediately have a hypersequent calculus for \textbf{\L} using \textbf{GA} which we refine to obtain a more direct and natural calculus for \textbf{\L}.

We start by interpreting hypersequents for \textbf{\L}. We extend the standard notion of validity for \textbf{\L} from $\models_\textrm{\L}$ to $\models^*_\textrm{\L}$ using the characteristic model $[-1,0]_\textrm{\L}$. A formula is interpreted as usual for this model but multisets of formulae are interpreted (as for \textbf{A}) as \emph{sums} of elements in $\mathbb{Q}$ (as opposed to bounded sums using the \L ukasiewicz conjunction $\oplus$). Interpretations are extended to sequents and hypersequents in the usual way.
\begin{definition}[Interpretation of Hypersequents for \L] \label{inthypl}
$\models^*_\textrm{\L} \Gamma_1 \vdash \Delta_1|\ldots|\Gamma_n \vdash \Delta_n$ iff for all $v:For \rightarrow [-1,0]_\textrm{\L}$ there exists $i$ such that $\Sigma_{A \in 
\Gamma_i} v(A) \le \Sigma_{B \in \Delta_i} 
v(B)$.
\end{definition}
We emphasise that for \emph{formulae} this notion coincides with the standard notion of validity for \textbf{\L}.
\begin{proposition}
$\models^*_\textrm{\L} \phi$ iff $\models_\textrm{\L} \phi$.
\end{proposition}
\textbf{Proof.} Immediate from Definition \ref{inthypl}. $\Box$\\[.1in]
Clearly our hypersequents for \textbf{\L} can also be interpreted in \textbf{A}. The following translation, extending that given by Definition \ref{deftrans} makes this relationship precise.
\begin{definition}[Translation *] \label{defexttrans}
$p^* = (p \lor q^\bot) \land t$, $\bot^* = q^\bot \land t$, $(A \limp B)^* = t \land (A^* \to B^*)$, $(\Gamma 
\vdash \Delta)^* = \Gamma^* \vdash \Delta^*$, $(S_1|\ldots|S_n)^* = 
S_1^*|\ldots|S_n^*$ where $q^\bot$ is an arbitary fixed propositional variable.
\end{definition}
\begin{theorem} \label{thmexttrans}
$\models^*_\textrm{\L} G$ iff $\models_A G^*$.
\end{theorem}
\textbf{Proof.} The proof of Theorem \ref{thmtrans} is easily extended from formulae to hypersequents. $\Box$\\[.1in]
Theorem \ref{thmexttrans} means that we have a calculus for \textbf{\L} simply by using \textbf{GA} applied to formulae of the form $A^*$. We refine this calculus in several steps. We first simplify the rules for formulae of the form $A^*$ to obtain the following calculus:
\begin{definition}[$\mathbf{GA^*}$]
$\mathbf{GA^*}$ consists of the axioms and structural rules of \textbf{GA} together with the following logical rules:
\[\small{
\newcommand*{\limprightrule}{\begin{array}[t]{c}
G|\Gamma,A^* \vdash B^*,\Delta \ \ G|\Gamma \vdash \Delta\\
\hline
G|\Gamma \vdash (A \limp B)^*, \Delta
\end{array}}
\newcommand*{\limpleftrule}{\begin{array}[t]{c}
G|\Gamma,B^* \vdash A^*,\Delta|\Gamma \vdash \Delta\\
\hline
G|\Gamma, (A \limp B)^* \vdash \Delta
\end{array}}
\newcommand*{\qstarrightrule}{\begin{array}[t]{c}
G|\Gamma \vdash \Delta \ \ G|\Gamma \vdash \Delta,q|\Gamma \vdash q^\bot,\Delta\\
\hline
G|\Gamma \vdash q^*, \Delta
\end{array}}
\newcommand*{\qstarleftrule}{\begin{array}[t]{c}
G|\Gamma \vdash \Delta|\Gamma,q^\bot \vdash \Delta \ \ G|\Gamma \vdash \Delta|\Gamma,q \vdash \Delta\\
\hline
G|\Gamma, q^* \vdash \Delta
\end{array}}
\newcommand*{\falstarrightrule}{\begin{array}[t]{c}
G|\Gamma \vdash \Delta \ \ G|\Gamma \vdash q^\bot,\Delta\\
\hline
G|\Gamma \vdash \bot^*, \Delta
\end{array}}
\newcommand*{\falstarleftrule}{\begin{array}[t]{c}
G|\Gamma \vdash \Delta|\Gamma,q^\bot \vdash \Delta\\
\hline
G|\Gamma, \bot^* \vdash \Delta
\end{array}}
\begin{array}[t]{lclc}
(\bot^*,l) & \falstarleftrule & (\bot^*,r) & \falstarrightrule\\
& & &\\
(q^*,l) & \qstarleftrule & (q^*,r) & \qstarrightrule\\ 
& & &\\
(\limp^*,l) & \limpleftrule & (\limp^*,r) & \limprightrule
\end{array}}
\]
\end{definition}
\begin{proposition} \label{propgastar}
If $G^*$ succeeds in $\textbf{GA}$ then $G^*$ succeeds in $\mathbf{GA^*}$.
\end{proposition}
\textbf{Proof.} It is easy to check that the logical rules of $\mathbf{GA^*}$ are derivable using the logical rules of \textbf{GA} and are hence invertible. Since they also reduce a hypersequent $G^*$ to atomic hypersequents and the two calculi share the same structural rules and axioms we are done. $\Box$\\[.1in]
Our next move is to define a calculus that is \emph{stronger} than $\mathbf{GA^*}$ and operates directly on formulae of \textbf{\L} rather than their translations:
\begin{definition}[G\L$\mathbf{_i}$] \textbf{G\L}$\mathbf{_i}$ consists of the axioms and structural rules of \textbf{GA} and the following logical rules:
\[
\small{
\newcommand*{\limprightrule}{\begin{array}[t]{c}
G|\Gamma,A \vdash B,\Delta \ \ G|\Gamma \vdash \Delta\\
\hline
G|\Gamma \vdash A \limp B, \Delta
\end{array}}
\newcommand*{\limpleftrule}{\begin{array}[t]{c}
G|\Gamma,B \vdash A,\Delta|\Gamma \vdash \Delta\\
\hline
G|\Gamma, A \limp B \vdash \Delta
\end{array}}
\newcommand*{\qstarrightrule}{\begin{array}[t]{c}
G|\Gamma \vdash q,\Delta|\Gamma \vdash \bot,\Delta\\
\hline
G|\Gamma \vdash q, \Delta
\end{array}}
\newcommand*{\qstarleftrule}{\begin{array}[t]{c}
G|\Gamma \vdash \Delta|\Gamma,q \vdash \Delta\\
\hline
G|\Gamma, q \vdash \Delta
\end{array}}
\newcommand*{\falstarleftrule}{\begin{array}[t]{c}
G|\Gamma \vdash \Delta|\Gamma,\bot \vdash \Delta\\
\hline
G|\Gamma, \bot \vdash \Delta
\end{array}}
\begin{array}[t]{lclc}
(\bot,l) & \falstarleftrule & &\\
& & &\\
(q,l) & \qstarleftrule & (q,r) & \qstarrightrule\\ 
& & &\\
(\limp,l) & \limpleftrule & (\limp,r) & \limprightrule
\end{array}
}\]
\end{definition}
\begin{proposition} \label{propli}
If $G^*$ succeeds in $\mathbf{GA^*}$ then $G$ succeeds in \textbf{G\L}$\mathbf{_i}$.
\end{proposition}
\textbf{Proof.} First we observe that removing the left premises from $(\bot^*,r)$ and $(q^*,l)$ and $(q^*,r)$ gives a calculus $\mathbf{GA_1^*}$ such that if $G^*$ succeeds in $\mathbf{GA^*}$ then $G^*$ succeeds in $\mathbf{GA^*_1}$. 
We now replace (for the calculus $\mathbf{GA^*_1}$) $q$ by $q^*$ in the rules $(q^*,l)$ and $(q^*,r)$ and $q^\bot$ by $\bot^*$ in the rules $(\bot^*,l)$ and $(\bot^*,r)$ to obtain a calculus $\mathbf{GA^*_2}$. Since no rules have conclusions with formulae that must be atomic we have that if $G^*$ succeeds in $\mathbf{GA^*_1}$ then $G^*$ succeeds in $\mathbf{GA^*_2}$. But this is just the translation of the calculus \textbf{G\L}$\mathbf{_i}$ (removing $(\bot^*,r)$ as it is just the trivial rule) so we have that if $G^*$ succeeds in $\mathbf{GA^*_2}$ then $G$ succeeds in \textbf{G\L}$\mathbf{_i}$ and we are done. $\Box$\\[.1in]
We now present our final calculus for \textbf{\L}:
\begin{definition}[G\L]
\textbf{G\L} has the following rules:\\[.1in]
\small{
Axioms
\[
\begin{array}[t]{lcclcclc}
(ID) & A \vdash A & \ \ \ \ \ \ & (\Lambda) & \vdash & \ \ \ \ \ \ \ & (\bot) & \bot \vdash 
A
\end{array}\]
Structural rules
\[
\newcommand*{\extweak}{\begin{array}[t]{c}
G|\Gamma \vdash \Delta\\
\hline
G|\Gamma \vdash \Delta|\Gamma' \vdash \Delta'
\end{array}}
\newcommand*{\extcont}{\begin{array}[t]{c}
G|\Gamma \vdash \Delta|\Gamma \vdash \Delta\\
\hline
G|\Gamma \vdash \Delta
\end{array}}
\newcommand*{\extsc}{\begin{array}[t]{c}
G|\Gamma_1,\Gamma_2 \vdash \Delta_1,\Delta_2\\
\hline
G|\Gamma_1 \vdash \Delta_2|\Gamma_2 \vdash \Delta_2
\end{array}}
\newcommand*{\extmingle}{\begin{array}[t]{c}
G|\Gamma_1 \vdash \Delta_1 \ \ G|\Gamma_2 \vdash \Delta_2\\
\hline
G|\Gamma_1,\Gamma_2 \vdash \Delta_1,\Delta_2
\end{array}}
\newcommand*{\cutrule}{\begin{array}[t]{c}
G|\Gamma, A \vdash \Delta,A\\
\hline
G|\Gamma \vdash \Delta
\end{array}}
\newcommand*{\intweak}{\begin{array}[t]{c}
G|\Gamma \vdash \Delta\\
\hline
G|\Gamma,A \vdash \Delta
\end{array}}
\begin{array}[t]{lclc}
(IW) & \intweak & &\\
& & &\\
(EW) & \extweak & (EC) & \extcont\\
& & &\\
(S) & \extsc & (M) & \extmingle
\end{array}
\]
Logical rules
\[
\newcommand*{\limprightrule}{\begin{array}[t]{c}
G|\Gamma,A \vdash B,\Delta \ \ G|\Gamma \vdash \Delta\\
\hline
G|\Gamma \vdash A \limp B, \Delta
\end{array}}
\newcommand*{\limpleftrule}{\begin{array}[t]{c}
G|\Gamma,B \vdash A,\Delta|\Gamma \vdash \Delta\\
\hline
G|\Gamma, A \limp B \vdash \Delta
\end{array}}
\newcommand*{\andrightrule}{\begin{array}[t]{c}
G|\Gamma \vdash A,\Delta \ \ G|\Gamma \vdash B,\Delta\\
\hline
G|\Gamma \vdash A \land B, \Delta
\end{array}}
\newcommand*{\andleftrule}{\begin{array}[t]{c}
G|\Gamma,A \vdash \Delta\\
\hline
G|\Gamma, A \land B \vdash \Delta
\end{array}}
\newcommand*{\orleftrule}{\begin{array}[t]{c}
G|\Gamma, A \vdash \Delta \ \ G|\Gamma, B \vdash \Delta\\
\hline
G|\Gamma, A \lor B \vdash \Delta
\end{array}}
\newcommand*{\orrightrule}{\begin{array}[t]{c}
G|\Gamma \vdash A, \Delta\\
\hline
G|\Gamma \vdash A \lor B \vdash \Delta
\end{array}}
\newcommand*{\neworrightrule}{\begin{array}[t]{c}
G|\Gamma \vdash A, \Delta|\Gamma \vdash B,\Delta\\
\hline
G|\Gamma \vdash A \lor B, \Delta
\end{array}}
\newcommand*{\newandleftrule}{\begin{array}[t]{c}
G|\Gamma,A \vdash \Delta|\Gamma,B \vdash \Delta\\
\hline
G|\Gamma, A \land B \vdash \Delta
\end{array}}
\begin{array}[t]{lclc}
(\limp,l) & \limpleftrule & (\limp,r) & \limprightrule
\end{array}
\]}
\end{definition}
Below we give a proof of the characteristic axiom of \textbf{\L}.
\begin{example} \label{examb}
A proof of $((A \limp B) \limp B) \limp ((B \limp A) \limp A)$ in 
\textbf{G\L}:
\small{
\[\infer{\vdash ((A \limp B) \limp B) \limp ((B \limp A) \limp A)}{
	\vdash &
	\infer{(A \limp B) \limp B \vdash (B \limp A) \limp A}{
		\infer{(A \limp B) \limp B \vdash}{
			\vdash} &
		\infer{(A \limp B) \limp B, B \limp A \vdash A}{
			\infer{B, B \limp A \vdash A, A \limp B|B \limp A \vdash 
A}{
				\infer{B, B \limp A \vdash A, A \limp B}{
					\infer{B,B \limp A \vdash A}{
						\infer{B,A \vdash A,B|B \vdash 
A}{
							\infer{B,A \vdash A,B}{
								B \vdash B & A 
\vdash A}}} &
					\infer{B,B \limp A,A \vdash A,B}{
						\infer{B,A \vdash A,B}{
							B \vdash B & A \vdash 
A}}}}}}}
\]}
\end{example}
We prove soundness and completeness proof-theoretically using the translation of Definition \ref{defexttrans}.
\begin{proposition} \label{propderiv}
The translations of the rules 
$(IW)$, $(\bot)$, $(\limp,l)$ and $(\limp,r)$ are derivable in \textbf{GA} for hypersequents containing only formulae of the form $A^*$.
\end{proposition}
\textbf{Proof.} For $(IW)$ we note that all formulae $A^*$ are of the form $X \land t$ and proceed as follows:
\[
\infer{G|\Gamma,X \land t \vdash \Delta}{
	\infer{G|\Gamma,X \vdash \Delta|\Gamma,t \vdash \Delta}{
		\infer{G|\Gamma,t \vdash \Delta}{
			G|\Gamma \vdash \Delta}}}
\]
For $(\bot)$ we proceed by induction on the complexity of $A$. If $A$ is a propositional variable $p$ then we have:
\[
\infer{t \land q^\bot \vdash (p \lor q^\bot) \land t}{
	\infer{t \land q^\bot \vdash p \lor q^\bot}{
		\infer{t \vdash p \lor q^\bot|q^\bot \vdash p \lor q^\bot}{
			\infer{q^\bot \vdash p \lor q^\bot}{
				\infer{q^\bot \vdash p|q^\bot \vdash q^\bot}{
					q^\bot \vdash q^\bot}}}} &
	\infer{t \land q^\bot \vdash t}{
		\infer{t \vdash t|q^\bot \vdash t}{
			t \vdash t}}}
\]
If $A = \bot$ then we succeed immediately by $(ID)$. For $A = B \limp C$ we have the following situation:
\[
\infer{t \land q^\bot \vdash t \land (B^* \to C^*)}{
	\infer{t \land q^\bot \vdash t}{
		\infer{t \vdash t|q^\bot \vdash t}{
			t \vdash t}} &
	\infer{t \land q^\bot \vdash B^* \to C^*}{
		t \land q^\bot, B^* \vdash C^*}}
\]
But now since $(IW)$ is derivable in \textbf{GA} we can step to $t \land q^\bot \vdash C^*$ for the right branch which succeeds by the induction hypothesis.\\[.1in]
$(\limp,l)$ and $(\limp,r)$ are derived as follows:
\[
\infer{G|\Gamma, t \land (A^* \to B^*) \vdash \Delta}{
	\infer{G|\Gamma,t \vdash \Delta|\Gamma, A^* \to B^* \vdash \Delta}{
		\infer{G|\Gamma \vdash \Delta|\Gamma, A^* \to B^* \vdash \Delta}{
			G|\Gamma \vdash \Delta|\Gamma,B^* \vdash \Delta,A^*}}}
\]
\[
\infer{G|\Gamma \vdash \Delta, t \land (A^* \to B^*)}{
	\infer{G|\Gamma \vdash \Delta,t}{
		G|\Gamma \vdash \Delta} &
	\infer{G|\Gamma \vdash \Delta, A^* \to B^*}{
		G|\Gamma,A^* \vdash \Delta, B^*}}
\Box
\]
\begin{proposition} \label{propbot}
The following rule is admissible in \textbf{G\L}:
\[
\infer{G|\Gamma_1 \vdash A, \Delta_1|\ldots|\Gamma_n \vdash A, \Delta_n}{
	G|\Gamma_1 \vdash \bot, \Delta_1|\ldots|\Gamma_n \vdash \bot, \Delta_n}
\]
\end{proposition}
\textbf{Proof.} By induction on the height $h$ of a proof of $G|\Gamma_1 \vdash \bot, \Delta_1|\ldots|\Gamma_n \vdash \bot, \Delta_n$. If $h=0$ then we must have $\bot \vdash \bot$ and we get $\bot \vdash A$ by $(\bot)$. The cases for $h>0$ follow easily from the induction hypothesis. $\Box$.
\begin{proposition} \label{propadmis}
The rules $(\bot,l)$, $(q,l)$ and $(q,r)$ of \textbf{G\L}$\mathbf{_i}$ are admissible in \textbf{G\L}.
\end{proposition}
\textbf{Proof.} $(q,l)$ is derivable using $(EC)$ and $(IW)$. $(q,r)$ and $(\bot,l)$ are admissible using $(EC)$ and Proposition \ref{propbot}.
\begin{proposition} \label{propgiffgstar}
$G$ succeeds in \textbf{G\L} iff $G^*$ succeeds in \textbf{GA}.
\end{proposition}
\textbf{Proof.}\\[.1in]
For the left-to-right direction we simply note that by Proposition \ref{propderiv} we have that the translations of all the rules of \textbf{G\L} are either rules of \textbf{GA} or derivable in \textbf{GA} for hypersequents containing only formulae of the form $A^*$.\\[.1in]
For the right-to-left direction by Propositions \ref{propli} and \ref{propgastar} we only have to show that if $G$ succeeds in \textbf{G\L}$\mathbf{_i}$ then $G$ succeeds in \textbf{G\L}. We simply note that all the extra rules of \textbf{G\L}$\mathbf{_i}$ are either derivable or admissible in \textbf{G\L} by Proposition \ref{propadmis}. $\Box$
\begin{theorem}[Soundness and Completeness of G\L]
$G$ succeeds in \textbf{G\L} iff $\models^*_\textbf{\L} G$.
\end{theorem}
\textbf{Proof.} By Proposition \ref{propgiffgstar}, Theorems \ref{thmcompga} and \ref{thmsoundga}, and Theorem \ref{thmexttrans} respectively we have that $G$ succeeds in \textbf{G\L} iff $G^*$ succeeds in \textbf{GA} iff $\models_A G^*$ iff $\models^*_\textrm{\L} G$. $\Box$
\begin{corollary}
$\vdash \phi$ succeeds in \textbf{G\L} iff $\models_\textrm{\L} \phi$.
\end{corollary}
Finally we note that by disregarding the rule $(\bot)$ we obtain a calculus for the positive part of \L ukasiewicz logic \textbf{\L}$\mathbf{^+}$. Moreover unlike \textbf{\L} we can interpret hypersequents for \textbf{\L}$\mathbf{^+}$ within \textbf{\L}$\mathbf{^+}$ in the standard way using $\oplus$ instead of $+$.
\section{Terminating Hypersequent Calculi}
Although \textbf{GA} and \textbf{G\L} are natural and elegant calculi they 
are not particularly suitable for proof search. Most obviously they are 
\emph{non-terminating} due to the presence of the external contraction rule 
$(EC)$. One method of obtaining terminating calculi is to replace $(EC)$ and 
$(S)$ with new rules that apply external contraction and splitting more 
carefully. In particular propositional variables can be removed from  
hypersequents \emph{one at a time} by adding a particular propositional variable as a \emph{focus} to each hypersequent; the focus is changed only when all occurrences of the marked propositional variable have been 
removed from the hypersequent.
\begin{definition}[Focussed Hypersequent]
If $G$ is a hypersequent and $q$ is a propositional variable then $[q]G$ is a 
focussed hypersequent with focus $q$. The interpretation of $[q]G$ is the same as for $G$ 
ie $\models_X [q]G$ iff $\models_X G$.
\end{definition}
\subsection{A Terminating Hypersequent Calculus for A}
We present the following terminating hypersequent calculus for \textbf{A}.
\begin{definition}[$\mathbf{GA_t}$]
$\mathbf{GA_t}$ consists of the axioms, logical rules and structural rules $(EW)$ and $(M)$ of \textbf{GA} with focussed hypersequents and the same focus for premises and conclusion, and also:
\[
\small{
\newcommand*{\extreduce}{\begin{array}[t]{c}
[p]G|m\Gamma_1,n\Gamma_2 \vdash m\Delta_1,n\Delta_2|S\\
\hline
[p]G|\Gamma_1, np \vdash \Delta_1|\Gamma_2 \vdash \Delta_2,mp
\end{array}}
\newcommand*{\cond}{\begin{array}[t]{l}
\Gamma_1, \Gamma_2, \Delta_1, \Delta_2 \textrm{ atomic}\\
n > 0, m > 0\\
p \not \in \Gamma_1 \cup \Delta_1 \cup \Gamma_2 \cup \Delta_2\\
S \textrm{ is } \Gamma_1,np \vdash \Delta_1 \textrm{ or } \Gamma_2 \vdash 
\Delta_2,mp
\end{array}}
\newcommand*{\condb}{\begin{array}[t]{l}
\textrm{$q$ occurs in $G$}\\
\textrm{$p$ doesn't occur in $G$}
\end{array}}
\newcommand*{\extshift}{\begin{array}[t]{c}
[q]G\\
\hline
[p]G
\end{array}}
\begin{array}[t]{lcll}
(shift) & \extshift & \textrm{where:} & \condb\\
& & &\\
(S') & \extreduce & \textrm{where:} & \cond
\end{array}}
\]
\end{definition}
Consider the example below, noting that some of the more obvious steps in the proof have been left out to aid readability.
\begin{example} A proof of $((q+q+q) \land (p+p+p)) \to (p+q+q)$ in $\mathbf{GA_t}$:
\[
\infer{[q]\vdash ((q+q+q) \land (p+p+p)) \to (p+q+q)}{
	\infer{[q](q+q+q) \land (p+p+p) \vdash p+q+q}{
		\infer{[q](q+q+q) \land (p+p+p) \vdash p,q,q}{
			\infer{[q]q+q+q \vdash p,q,q|p+p+p \vdash p,q,q}{
				\infer{[q]q,q,q \vdash p,q,q|p,p,p \vdash p,q}{
					\infer{[q]q \vdash p|p,p \vdash q,q \ \ \ [1]}{
						\infer{[q]p,p \vdash p,p|q \vdash p}{
						\infer{[q]p,p \vdash p,p}{
						[q]p \vdash p & [q]p \vdash p}}}}}}}}
\]
[1] For the application of $(S')$ at this point in the proof we have $n=1$ occurrences of the focus $q$ on the left hand side of the first component and $m=2$ occurrences of $q$ on the right hand side of the second component. Hence for the next line up in the proof we have a component with $2$ copies of the first component put together with $1$ copy of the second component.
\end{example}
Let us see why this calculus terminates. 
Clearly the logical rules serve to reduce any hypersequent to a hypersequent 
containing only atoms. The proof then \emph{focuses} on a particular 
propositional variable, $p$ say, and uses $(S')$, $(M)$, $(ID)$ and $(EW)$ to 
remove all occurrences of $p$. Each 
component has a difference between the number of $ps$ on each side; $(S')$ 
reduces the sum of differences for 
the whole hypersequent. When this sum is 0 only $(M)$, $(ID)$ and $(EW)$ can be used 
to remove further occurrences of 
$p$. If a propositional variable has been removed from the hypersequent 
completely the $(shift)$ rule allows a 
change of focus to a different variable. 

We now put this more formally.
\begin{definition}[count]
$count(\Gamma,p) = |p : p \in \Gamma|$
\end{definition}
\begin{definition}[$\mathbf{d([p]G)}$]
For a hypersequent $G=\Gamma_1 \vdash \Delta_1|\ldots|\Gamma_n \vdash \Delta_n$, 
 $d([p]G)=\Sigma_{i=1}^n 
|count(\Gamma_i,p) - count(\Delta_i,p)|$.
\end{definition}
\begin{proposition} \label{propdred}
$d([p]G|m\Gamma_1,n\Gamma_2 \vdash m\Delta_1,n\Delta_2|S) < d([p]G|\Gamma_1, np 
\vdash \Delta_1|\Gamma_2 \vdash 
\Delta_2,mp)$ when $\Gamma_1, \Gamma_2, \Delta_1, \Delta_2$ are atomic, $n>0$, 
$m>0$, 
$p \not \in \Gamma_1 \cup \Delta_1 \cup \Gamma_2 \cup \Delta_2$ and $S$ is 
$\Gamma_1,np \vdash \Delta_1$ or $\Gamma_2 \vdash 
\Delta_2,mp$.
\end{proposition}
\textbf{Proof.} Suppose that $S = \Gamma_1,np \vdash \Delta_1$ then 
$d([p]G|m\Gamma_1,n\Gamma_2 \vdash m\Delta_1,n\Delta_2|S) = d([p]G) + n < d([p]G) 
+ n + m = d([p]G|\Gamma_1, np \vdash \Delta_1|\Gamma_2 \vdash 
\Delta_2,mp)$. The case where $S = \Gamma_2 \vdash \Delta_2,mp$ is symmetrical. 
$\Box$
\begin{theorem}[Termination] \label{thmgatterm}
$\mathbf{GA_t}$ terminates for $[p]G$.
\end{theorem}
\textbf{Proof.} We proceed by induction on $(c,n,d,s)$ ordered lexicographically 
where $c$ is the multiset complexity of $G$ not including atoms, $n$ is the 
number of different 
propositional variables in $[p]G$ (including $p$), $d$ is $d([p]G)$ and $s$ is 
the number of symbols in $G$. 
We show that all the rules (read backwards) strictly \emph{decrease} 
$(c,n,d,s)$. The logical rules all 
strictly reduce $c$ so we turn our attention to the structural rules. $(shift)$ 
does not increase $c$ and 
decreases $n$. $(S')$ does not increase $c$ or 
$n$ and by 
Proposition \ref{propdred} strictly reduces $d$. $(EW)$ and $(M)$ do not increase $c$, $n$ or 
$d$ and strictly 
reduce $s$. $\Box$\\[.1in]
We now check that $\mathbf{GA_t}$ is sound and complete.
\begin{theorem}[Soundness]
If $[p]G$ succeeds in $\mathbf{GA_t}$ then $\models_\mathbf{A} G$.
\end{theorem}
\textbf{Proof.} We observe simply that $(S')$ is a derived rule of \textbf{GA}. 
$\Box$\\[.1in]
Although $(S')$ is not invertible we have that if its conclusion is valid then its premise is valid for \emph{one of the choices for S}.
\begin{proposition} \label{propsemiinv}
If $\models_A G|\Gamma_1 \vdash \Delta_1 | \Gamma_2 \vdash \Delta_2$ then either 
$\models_A 
G|\Gamma_1,\Gamma_2 \vdash \Delta_1, \Delta_2|\Gamma_1 \vdash \Delta_1$ or 
$\models_A G|\Gamma_1,\Gamma_2 
\vdash \Delta_1, \Delta_2|\Gamma_2 \vdash \Delta_2$.
\end{proposition}
\textbf{Proof.} Let $G = \Gamma'_1 \vdash \Delta'_1|\ldots|\Gamma'_n \vdash 
\Delta'_n$. By Proposition \ref{proplin} we have 
that $\models_A G|\Gamma_1 \vdash \Delta_1 | \Gamma_2 \vdash \Delta_2$ iff there 
exist $\lambda_1, 
\lambda_2, \mu_1, \ldots, \mu_n$ where $\lambda_i > 0$ for $1 \le i \le 2$ or $\mu_i > 0$ for $1 \le i \le n$, such that $\Sigma_{i=1}^n \mu_i \Gamma'_i + 
\lambda_1 \Gamma_1 + 
\lambda_2 \Gamma_2 = \Sigma_{i=1}^n \mu_i \Delta'_i + \lambda_1 \Delta_1 + 
\lambda_2 \Delta_2$. Suppose  that $\lambda_1 \le \lambda_2$ then we have $\Sigma_{i=1}^n \mu_i \Gamma'_i + \lambda_1 (\Gamma_1 + 
\Gamma_2) + (\lambda_2 - \lambda_1) \Gamma_2 = \Sigma_{i=1}^n \mu_i \Delta'_i + \lambda_1 (\Delta_1 + \Delta_2) + (\lambda_2 - \lambda_1) \Delta_2$ which gives, again by Proposition \ref{proplin}, that $\models_A G|\Gamma_1,\Gamma_2 \vdash \Delta_1, \Delta_2|\Gamma_2 \vdash \Delta_2$. The case where $\lambda_1 \ge \lambda_2$ is symmetrical. $\Box$\\[.1in]
Now consider a valid hypersequent $G$ where for a given propositional variable $p$ there are never more occurrences of $p$ on the right hand side of a component of $G$ than the left. If we remove a component where $p$ occurs \emph{more} on the left than the right then we will still obtain a valid hypersequent $G'$.
\begin{proposition} \label{propremacom}
Given an atomic hypersequent $G = \Gamma_1 \vdash \Delta_1| \ldots |\Gamma_n \vdash \Delta_n$ where $\models_A G$ 
and $count(\Gamma_1,p) > count(\Delta_1,p)$ and $count(\Gamma_i,p) \ge 
count(\Delta_i,p)$ for $i=2 \ldots 
n$ then $\models_A \Gamma_2 \vdash \Delta_2|\ldots|\Gamma_n \vdash \Delta_n$.
\end{proposition}
\textbf{Proof.} By Proposition \ref{proplin} $\models_A G$ iff there exist 
$\lambda_1,\ldots,\lambda_n \in \mathbb{Z}^+$ such 
that $\lambda_i > 0$ for some $i$, $1 \le i \le n$ and $\cup_{i=1}^n \lambda_i 
\Gamma_i = \cup_{i=1}^n \lambda_i \Delta_i$. But if $\lambda_1 > 0$ then 
$count(\cup_{i=1}^n \lambda_i 
\Gamma_i,p) > count(\cup_{i=1}^n \lambda_i \Delta_i,p)$ a contradiction. Hence 
$\lambda_1 = 0$ and we 
have $\models_A \Gamma_2 \vdash \Delta_2|\ldots|\Gamma_n \vdash \Delta_n$. $\Box$\\[.1in]
Our completeness proof proceeds along the same lines as that of Theorem \ref{thmcompga}.
\begin{theorem}[Completeness] \label{thmcompgt}
If $\models_\mathrm{A} G$ then $[p]G$ succeeds in $\mathbf{GA_t}$.
\end{theorem}
\textbf{Proof.} Suppose 
$\models_\mathbf{A} G$ then 
by applying the (invertible) logical rules to $[p]G$ we get valid 
atomic hypersequents. It 
remains to show then that all valid atomic hypersequent $[p]G'$ are provable. We 
proceed by induction on $(n,d)$ where $n$ is the number of propositional 
variables in $G'$ and $d$ is 
$d([p]G')$. If $n=0$ then we just apply $(EW)$ and 
$(\Lambda)$ and we are done. For $n>0$ if $d=0$ then we apply $(M)$ and $(ID)$ 
until there are no 
occurrences of $p$ left in the hypersequent and use $(shift)$ to decrease $n$. 
If $d>0$ then for 
some component $\Gamma_1 \vdash \Delta_1$ either $count(\Gamma_1,p) > 
count(\Delta_1,p)$ or 
$count(\Delta_1,p) > count(\Gamma_1,p)$. Suppose the former; if there is no 
component $\Gamma_2 
\vdash \Delta_2$ in $[p]G'$ where $count(\Delta_2,p) > count(\Gamma_2,p)$ then 
by Proposition \ref{propremacom} we can 
apply $(EW)$ to $[p]G'$ to remove $\Gamma_1 \vdash \Delta_1$ giving $[p]G''$ 
where $\models_A [p]G''$ and 
$d([p]G'') < d([p]G')$. Now suppose there is such a component $\Gamma_2 
\vdash \Delta_2$ in $[p]G'$ where $count(\Delta_2,p) > count(\Gamma_2,p)$. We 
apply 
$(M)$ and $(ID)$ to $[p]G'$ giving $[p]G''$ containing components $\Gamma_1',np 
\vdash \Delta_1'$ and 
$\Gamma_2' \vdash \Delta_2',mp$ where $n>0$, $m>0$ and $p \not \in 
\Gamma'_1,\Gamma'_2,\Delta'_1,\Delta'_2$. So by Proposition \ref{propsemiinv} we can apply $(S')$ to obtain a hypersequent $[p]G'''$ where $\models_A [p]G'''$ and by Proposition \ref{propdred} $d([p]G''') < d([p]G'') \le 
d([p]G')$. 
$\Box$
\subsection{A Terminating Hypersequent Calculus for \L}
A terminating hypersequent calculus for \textbf{\L} is developed in a similar way to $\mathbf{GA_t}$.
\begin{definition}[$\mathbf{G\textbf{\L}_t}$]
$\mathbf{G\textbf{\L}_t}$ consists of the axioms, logical rules and structural rules $(IW)$, $(EW)$ and $(M)$ of \textbf{G\L} with foccussed hypersequents 
and the same focus for premises and conclusion, and also the rules $(shift)$ and $(S')$ of $\mathbf{GA_t}$.
\end{definition}
\begin{theorem}[Termination of $\mathbf{G\textbf{\L}_t}$]
$\mathbf{G\textbf{\L}_t}$ is terminating. 
\end{theorem}
\textbf{Proof.} Similar to the proof of Theorem \ref{thmgatterm}. Just observe that $(IW)$ does not increase the multiset complexity of $G$ not including atoms, the number of different atomic variables or $d([p]G)$ and strictly reduces the number of symbols of $G$. $\Box$
\begin{theorem}[Soundness of $\textbf{G\L}\mathbf{_t}$]
If $[p]G$ succeeds in \textbf{G\L}$\mathbf{_t}$ then $\models^*_\textrm{\L} G$.
\end{theorem}
\textbf{Proof.} $(S')$ is a derived rule of \textbf{G\L} so we are done. $\Box$
Completeness for $\mathbf{G\textbf{\L}_t}$ is proved in the same way as for $\mathbf{GA_t}$ with the following proposition replacing Proposition \ref{propremacom}. Note that for convenience we call formulae of the form $q^*$ where $q$ is atomic, \emph{starred atoms}.
\begin{proposition} \label{propremanat}
Given a hypersequent $G = \Gamma_1 \vdash \Delta_1| \ldots |\Gamma_n \vdash \Delta_n$ containing only starred atoms where $\models_A G$ and $count(\Gamma_1,p^*) > count(\Delta_1,p^*)$ and $count(\Gamma_i,p^*) \ge count(\Delta_i,p^*)$ for $i=2 \ldots n$ then $\models_A \Gamma_1 - \{p^*\} \vdash \Delta_1|\Gamma_2 \vdash \Delta_2|\ldots|\Gamma_n \vdash \Delta_n$.
\end{proposition}
\textbf{Proof.} $G$ is valid iff the atomic hypersequents obtained by applying the $\land$ and $\lor$ rules of \textbf{GA} to the starred atoms of $G$ are valid. But these atomic hypersequents meet the conditions of Proposition \ref{propremacom} so we can remove the components containing a $p$ resulting directly from the component $\Gamma_1 \vdash \Delta_1$. The resulting hypersequents are valid and therefore so is the hypersequent $\models_A \Gamma_1 - \{p^*\} \vdash \Delta_1|\Gamma_2 \vdash \Delta_2|\ldots|\Gamma_n \vdash \Delta_n$.
\begin{theorem}[Completeness of $\textbf{G\L}\mathbf{_t}$]
If $\models^*_\textrm{\L} G$ then $[p]G$ succeeds in \textbf{G\L}$\mathbf{_t}$.
\end{theorem}
\textbf{Proof.} Let $\mathbf{GA_t^*}$ be the calculus \textbf{G\L}$\mathbf{_t}$ with formulae $A$ replaced everywhere with $A^*$ and the requirements for atoms in the structural rules changed to requirements for starred atoms. We have if $[p]G^*$ succeeds in $\mathbf{GA_t^*}$ then $[p]G$ succeeds in \textbf{G\L}$\mathbf{_t}$. Now suppose $\models^*_\mathrm{\L} G$, by Proposition \ref{thmexttrans}, we have $\models_A G^*$. $[p]G^*$ is shown to succeed in $\mathbf{GA_t^*}$ imitating the proof of Theorem \ref{thmcompgt}. We just note that Propositions \ref{propsemiinv} and \ref{propdred} hold when applied to starred atoms and that Proposition \ref{propremacom} can be replaced by Proposition \ref{propremanat}. $\Box$
\section{Labelled Single Sequent Calculi}
One drawback of our hypersequent calculi from a \emph{computational} perspective is that components multiply exponentially with respect to occurrences of $\lor$ and $\land$. Here we give a method for tackling this problem using \emph{labels}. We represent a disjunction of several 
unlabelled components (ie a hypersequent) as a single \emph{labelled} sequent and give rules that operate on all components \emph{simultaneously}. The labels themselves are built up from a unit label $1$ and atomic labels $x_1, x_2, \ldots$ and each labelled formula in a sequent consists of a label and a formula. Unlabelled components may be obtained from labelled sequents via labelling functions that map each label into the set $\{0,1\}$, 
removing formulae labelled with a $0$ from the sequent and leaving those labelled with a $1$. For example, the sequent $x:p, 1:q \vdash 1:p, x:q$ is mapped by a labelling 
function $f$ to $q \vdash p$ if $f(x)=0$ and to $p,q \vdash p,q$ if $f(x)=1$, 
the corresponding hypersequent being $q \vdash p|p,q \vdash p,q$.

The following definitions make these notions precise:
\begin{definition} [(Atomic) Labels, Labelled Formulae, Labelled Sequents]
The set of labels $Lab$ is generated from the set of atomic labels $\{x_i\}_{i \in \mathbb{N}}$ as follows: (1) 
$1 \in Lab$, (2) $x_i \in Lab$ for all $i \in \mathbb{N}$, (3) if $x \in Lab$ 
and $y \in 
Lab$ then $xy \in Lab$. A labelled formula is of the form $x:A$ where $x \in 
Lab$ and $A$ is a formula. A labelled sequent is a sequent $\Gamma \vdash 
\Delta$ where $\Gamma$ and $\Delta$ contain only labelled formulae.
\end{definition}
\begin{definition}[Labelling Function]
$f:Lab \to \{0,1\}$ is a labelling function iff: (1) $f(1) = 1$,
(2) $f(x_i) \in \{0,1\}$ for all $i \in \mathbb{N}$,
(3) $f(xy) = f(x).f(y)$. $f$ is extended to multisets of labelled formulae and labelled sequents by the conditions:
(4) $f(\Gamma) = \{A \ | \ x:A \in \Gamma$ and $f(x)=1\}$,
(5) $f(\Gamma \vdash \Delta) = f(\Gamma) \vdash f(\Delta)$.
\end{definition}
\begin{definition}[$\mathbf{\Gamma^l}$, $\mathbf{\Gamma^{ul}}$]
Given a multiset of unlabelled formulae $\Gamma$, $\Gamma^l=\{1:A \ | \ A \in 
\Gamma\}$. Given 
a multiset of labelled formulae $\Gamma$, $\Gamma^{ul} = \{A \ | \ x:A \in 
\Gamma\}$.
\end{definition}
\subsection{A Labelled Single Sequent Calculus for A}
We interpret a labelled sequent $S$ for \textbf{A} as the disjunction of all the unlabelled sequents obtained by applying labelling functions to $S$.
\begin{definition}[Interpretation of Labelled Sequents for A] \label{defintlabseq}
Let the interpretation of a labelled sequent $S$ be $\phi^S = \bigvee \{\phi^{f(S)} | f$ a labelling function$\}$ so that $\models_A S$ iff $\models_A \phi^S$.
\end{definition}
Observe that validity for labelled sequents coincides with the usual notion of validity $\models_A$ for sequents where all formulae are labelled with a $1$.
\begin{proposition}
For multisets $\Gamma, \Delta$ we have $\models_A \Gamma^l \vdash \Delta^l$ iff $\models_A \Gamma \vdash \Delta$.
\end{proposition}
\textbf{Proof.} Immediate from Definition \ref{defintlabseq}.\\[.1in]
We now present our labelled calculus for \textbf{A}. Note that recalling Proposition \ref{proplimpdef}, we choose $\limp$ as a primitive connective rather than $\land$ and $\lor$. Our reasons are firstly that the rules for $\limp$ are more uniform than those for $\land$ and $\lor$ (and looking ahead to Section 7 the rules for $\land$ and $\lor$ for the unlabelled calculus do not even obey the subformula property) and secondly that we want to exploit similarities with \textbf{\L}.
\begin{definition}[$\mathbf{GA_l}$]
$\mathbf{GA_l}$ has the following rules:
\[
\small{
\newcommand*{\trightrule}{\begin{array}[t]{c}
\Gamma \vdash \Delta\\
\hline
\Gamma \vdash x:t, \Delta
\end{array}}
\newcommand*{\tleftrule}{\begin{array}[t]{c}
\Gamma \vdash \Delta\\
\hline
\Gamma,x:t \vdash \Delta
\end{array}}
\newcommand*{\lplusleft}{\begin{array}[t]{c}
\Gamma, x:A, x:B \vdash \Delta\\
\hline
\Gamma, x:A+B \vdash \Delta
\end{array}}
\newcommand*{\lplusright}{\begin{array}[t]{c}
\Gamma \vdash \Delta, x:A,x:B\\
\hline
\Gamma \vdash \Delta, x:A+B
\end{array}}
\newcommand*{\ltoleft}{\begin{array}[t]{c}
\Gamma,x:B \vdash \Delta,x:A\\
\hline
\Gamma,x:A \to B \vdash \Delta
\end{array}}
\newcommand*{\ltoright}{\begin{array}[t]{c}
\Gamma,x:A \vdash \Delta,x:B\\
\hline
\Gamma \vdash \Delta, x:A \to B
\end{array}}
\newcommand*{\llimpleft}{\begin{array}[t]{c}
\Gamma,xy:B \vdash \Delta,xy:A\\
\hline
\Gamma,x:A \limp B \vdash \Delta
\end{array}}
\newcommand*{\llimpright}{\begin{array}[t]{c}
\Gamma,x:A \vdash \Delta,x:B \ \ \Gamma \vdash \Delta\\
\hline
\Gamma \vdash \Delta, x:A \limp B
\end{array}}
\newcommand*{\lwk}{\begin{array}[t]{c}
\Gamma \vdash \Delta\\
\hline
\Gamma, x:A \limp B \vdash \Delta
\end{array}}
\newcommand*{\lognotleftrule}{\begin{array}[t]{c}
\Gamma \vdash x:A, \Delta\\
\hline
\Gamma, x:\lnot A \vdash \Delta
\end{array}}
\newcommand*{\lognotrightrule}{\begin{array}[t]{c}
\Gamma, x:A \vdash \Delta\\
\hline
\Gamma \vdash x:\lnot A, \Delta
\end{array}}
\begin{array}[t]{lclc}
(t,l) & \tleftrule & (t,r) & \trightrule\\
& & &\\
(\lnot,l) & \lognotleftrule & (\lnot,r) & \lognotrightrule\\
& & &\\
(+,l) & \lplusleft & (+,r) & \lplusright\\
& & &\\
(\to,l) & \ltoleft & (\to,r) & \ltoright\\
& & &\\
(\limp,l) & \llimpleft & (\limp,r) & \llimpright\\
& \textnormal{$y$ a new atomic label} & &
\end{array}}
\]
\[
\small{
\begin{array}[t]{lcl}
(success) & \Gamma \vdash \Delta & \textnormal{where $\Gamma$ and $\Delta$ are atomic and there exist labelling}\\
& & \textnormal{functions $f_1, \ldots f_n$ such that $\cup_{i=1}^n f_i(\Gamma) = \cup_{i=1}^n f_i(\Delta)$}
\end{array}}
\]
\end{definition}
Derived rules for $\land$ and $\lor$ are:
\[
\small{\newcommand*{\rulex}{\begin{array}[t]{c}
\Gamma, x:A, xy:B \vdash xy:A, \Delta\\
\hline
\Gamma, x:A \land B \vdash \Delta\\
\end{array}}
\newcommand*{\ruley}{\begin{array}[t]{c}
\Gamma \vdash \Delta, x:A \ \ \Gamma \vdash \Delta, x:B\\
\hline
\Gamma \vdash \Delta, x:A \land B\\
\end{array}}
\newcommand*{\rula}{\begin{array}[t]{c}
\Gamma, x:A \vdash \Delta \ \ \Gamma, x:B \vdash \Delta\\
\hline
\Gamma, x:A \lor B \vdash \Delta\\
\end{array}}
\newcommand*{\rulb}{\begin{array}[t]{c}
\Gamma, xy:A \vdash \Delta, x:A, xy:B\\
\hline
\Gamma \vdash \Delta, x:A \lor B\\
\end{array}}
\begin{array}[t]{lclc}
(\land,l) & \rulex & (\land,r) & \ruley\\
& & &\\
(\lor,l) & \rula & (\lor,r) & \rulb\\
\end{array}}\]
The following example shows that \emph{more than one} labelling function may be required to apply $(success)$:
\begin{example}
A proof of $(p \limp (p \limp r)) \limp ((q \limp (q \limp r)) \limp (p \limp (q \limp 
r)))$ in $\mathbf{GA_l}$ (note that for convenience we write $x$ instead of $x1$):
\[
\infer{\vdash 1:(p \limp (p \limp r)) \limp ((q \limp (q \limp r)) \limp (p \limp (q \limp 
r)))}{
	\vdash &
	\infer{1:p \limp (p \limp r) \vdash (q \limp (q \limp r)) \limp (p \limp (q \limp 
r))}{
	\infer{1:p \limp (p \limp r) \vdash}{
	\infer{u:p \limp r \vdash u:p}{
	\infer{uv:r \vdash u:p, uv:p}{
	f(u)=f(v)=0}}} &
	\infer{1:p \limp (p \limp r), 1:q \limp (q \limp r) \vdash p \limp (q \limp r)}{
	\infer{\vdots}{
	\infer{1:p \limp (p \limp r), 1:q \limp (q \limp r), 1:p , 1:q \vdash 1:r}{
	\infer{x:p \limp r, 1:q \limp (q \limp r), 1:p, 1:q \vdash 1:r, x:p}{
		\infer{xy:r, 1:q \limp (q \limp r), 1:p, 1:q \vdash 1:r, x:p,xy:p}{
			\infer{xy:r,z:q \limp r, 1:p, 1:q \vdash 1:r,x:p,xy:p,z:q}{
				xy:r,zw:r,1:p,1:q \vdash 1:r,x:p,xy:p,z:q,zw:q}}}}}}}}\]
For the right branch we apply $(success)$ with \emph{two} labelling functions:\\[.1in]
$f_1(x)=f_1(y)=1$, $f_1(z)=f_1(w)=0$\\
$f_2(x)=f_2(y)=0$, $f_2(z)=f_2(w)=1$\\[.1in]
Which gives:\\[.1in]
$\cup_{i=1}^2 f_i(\{xy:r,zw:r,1:p,1:q\}) = \cup_{j=1}^2 f_j(1:r,x:p,xy:p,z:q,zw:q) = 
\{r,r,p,p,q,q\}$
\end{example}
The $(success)$ rule replaces the structural rules of \textbf{GA}. Interpreting each labelling function applied to the labelled sequent as a component it can be seen that $(success)$ checks whether putting together any number of components (ie applying $(EC)$ and $(S)$) gives a component of the form $\Gamma \vdash \Gamma$ (which succeeds in \textbf{GA} using $(M)$ and $(ID)$). Note also that checking whether $(success)$ can be applied is equivalent to checking the inconsistency of the set of inequations $\{f(\Gamma) > f(\Delta) | f$ a labelling function$\}$ over $\mathbb{Q}$ ie to solving a \emph{linear programming problem}.

We now consider the soundness and completeness of $\mathbf{GA_l}$.
\begin{theorem}[Soundness of $\mathbf{GA_l}$] \label{thmsoundlga}
If $S$ succeeds in $\mathbf{GA_l}$ then $\models_A S$.
\end{theorem}
\textbf{Proof} We reason by induction on the length of a proof in 
$\mathbf{GA_l}$ and show that the 
logical rules and $(success)$ are sound in $\mathbb{Q}$.
\begin{itemize}
\item Logical rules. Consider $(\limp,l)$; if $\models_A \Gamma, xy:B \vdash 
\Delta, xy:A$ then 
given a valuation $v$ we have that $v(f(\Gamma))+v(f(xy:B)) \le v(f(\Delta)) + 
v(f(xy:A))$ for some 
labelling function $f$. For $f(xy)=0$ we have $v(f(\Gamma)) \le v(f(\Delta))$ so 
we take a labelling 
function $f'$ defined as $f'(z)=f(z)$ for $z \neq x$ and $f'(x)=0$ and get 
$v(f'(\Gamma)) + v(f'(x:A 
\limp B)) \le v(f'(\Delta))$. For $f(x)=f(y)=1$ we have $v(f(\Gamma)) + v(B) \le 
v(f(\Delta)) + v(A)$ so 
taking $f' = f$ we get $v(f'(\Gamma)) + v(f'(x:A \limp B)) \le v(f'(\Delta))$. 
Proofs for the other logical rules are very similar.
\item $(success)$. By $(success)$ there exist labelling functions $f_1, \ldots, 
f_n$ and $\lambda_1, 
\ldots, \lambda_n$ where $\lambda_i > 0$ for some $i$ $1 \le i \le n$ and 
$\cup_{i=1}^n 
\lambda_i.f_i(\Gamma) = \cup_{i=1}^n \lambda_i.f_i(\Delta)$ so by 
Proposition \ref{proplin} we have 
$\models_A f_1(\Gamma) \vdash f_1(\Delta)|\ldots|f_n(\Gamma) \vdash f_n(\Delta)$ 
as required. $\Box$
\end{itemize}  
\begin{proposition} \label{proplabinv}
The logical rules of $\mathbf{GA_l}$ are invertible.
\end{proposition}
\textbf{Proof.} We show the invertibility of $(\limp,l)$ as an example. If 
$\models_A \Gamma,x:A \limp B 
\vdash \Delta$ then given a valuation $v$ we have that $v(f(\Gamma)) + v(f(x:A 
\limp B)) \le 
v(f(\Delta))$ for some labelling function $f$. For $f(x)=0$ we take $f'=f$ and $f'(y)=0$ and get $v(f'(\Gamma)) + 
v(f'(xy:B)) \le 
v(f'(\Delta)) + v(f'(xy:A))$. For $f(x)=1$, if $v(A) \le v(B)$ then we have 
$v(f'(\Gamma)) \le v(f'(\Delta))$ 
 and we take $f'=f$ and $f'(y)=0$; if $v(A) > v(B)$ then we have $v(f'(\Gamma)) + 
v(B) \le v(f'(\Delta))+ 
v(B)$ and we take $f'=f$ and $f'(y)=1$. $\Box$
\begin{theorem}[Completeness of $\mathbf{GA_l}$] \label{thmcomplab}
If $\models_\mathbf{A} S$ then $S$ 
succeeds in $\mathbf{GA_l}$ .
\end{theorem}
\textbf{Proof} As in the proof of Theorem \ref{thmcompga} we apply the invertible (by Proposition \ref{proplabinv}) 
logical rules to $S$ obtaining a set of valid labelled atomic sequents. We now show 
that all valid labelled atomic sequents $\Gamma \vdash \Delta$ are provable. 
We have that there are 
labelling functions $f_1, \ldots, f_n$ such that $\models_A f_1(\Gamma) \vdash 
f_1(\Delta)|\ldots|f_n(\Gamma) \vdash f_n(\Delta)$; now by Proposition 
\ref{proplin} there exist 
$\lambda_1, \ldots, \lambda_n$ where $\lambda_i > 0$ for some 
$i$ $1 \le i \le n$ and $\cup_{i=1}^n \lambda_i.f_i(\Gamma) = \cup_{i=1}^n 
\lambda_i.f_i(\Delta)$ so we can apply $(success)$. $\Box$\\[.1in]
An apparent problem for the complexity of $\mathbf{GA_l}$ is that checking the $(success)$ rule is equivalent to 
checking the inconsistency of a set of equations of size exponential in the number of 
labels. As we show however, this set of equations can be transformed in linear time into a 
set of equations of linear size such that the new set is consistent iff the old set is 
consistent.

First however we introduce some new terminology.
\begin{definition}[Labelled Inequation] 
A labelled inequation is an inequation $\Gamma \triangleright \Delta$ where $\Gamma$ and $\Delta$ contain only labelled formulae, $\triangleright \in \{>,\ge\}$. For a labelling function $f$, $f(\Gamma \triangleright \Delta) = f(\Gamma) \triangleright f(\Delta)$, $\triangleright \in \{>,\ge\}$.
\end{definition}
\begin{definition}[Consistent Set of Labelled Inequations]
A set of labelled inequations $U$ is consistent iff the set $\{f(\Gamma) \triangleright f(\Delta) | \Gamma \triangleright \Delta \in U$, $f$ a labelling function$\}$ is consistent over $\mathbb{Q}$.
\end{definition}
\begin{definition}[Label-regular]
A labelled inequation is label-regular iff the set of atomic labels occurring in the inequation together with $1$ form a tree with root $1$ and each label occurring in the inequation is a path from $1$ to a 
node.
\end{definition}
\begin{definition}[Maximal Label]
Given a label-regular set of atomic labels, a maximal label is a child node of 
1 in the tree.
\end{definition}
The idea is that proofs in $\mathbf{GA_l}$ generate labelled sequents of a particular form. Each new atomic label $y$ is introduced in a proof as part of a complex label $xy$ and will always occurs alongside $x$ in subsequent labelled sequents in the proof.
\begin{proposition} \label{proplabreg}
Given a branch of a proof in $\mathbf{GA_l}$ $[\Gamma^l \vdash 
\Delta^l,\ldots,\Sigma \vdash \Pi]$ the labelled inequation $\Sigma > \Pi$ is 
label-regular.
\end{proposition}
\textbf{Proof.} By induction on the length of a proof in $\mathbf{GA_l}$. New 
labels are only introduced via $(\limp,l)$ and it is easy to see that the new atomic label $y$ can be added to the tree of atomic labels as a new node below those atomic labels occurring in $x$ and that the label 
$xy$ forms a path from the root to this node.
$\Box$\\[.1in]
We now prove our main lemma.
\begin{lemma}\label{lemredseq}
Given a set $U$ of $m$ label-regular inequations with $n$ atomic labels 
occurring in $U$ where no 
atomic label occurs in more than one inequation, a set $V$ of $2n+m$ unlabelled 
inequations can be found 
in $O(n)$ time such that $V$ is consistent iff $U$ is consistent.
\end{lemma}
\textbf{Proof.} We proceed by induction on $n$. For $n=0$ we take 
$V=\{\Gamma^{ul} \triangleright \Delta^{ul}|\Gamma \triangleright \Delta \in U\}$ and we are done. For $n>0$ there must be an inequation $S$ in $U$ containing a maximal atomic label $x$, where $S = \Gamma, \Sigma \triangleright \Delta, \Pi$, $\triangleright \in \{>,\ge\}$, $\Sigma$ and $\Pi$ contain only formulae in which $x$ occurs and $\Gamma$ and $\Delta$ contain only formulae in which $x$ does not occur. Now we define new inequations $S_1 = \Gamma, 1:\lambda_x \triangleright \Delta$, $S_2 = \Sigma \ge \Pi, 1:\lambda_x$ and $S3 = 0 \ge 1:\lambda_x$ where $\lambda_x$ is a new propositional variable and $U$ is consistent iff $U'=U - \{S\} \cup \{S_1,S_2,S_3\}$ with $n-1$ atomic labels is consistent. Now $S_1$, $S_2$ and $S_3$ are all label-regular and since $x$ is a maximal label there are no atomic labels occurring in more than one labelled inequation in $U'$. So by the induction hypothesis $U'$ is consistent iff $V$ is consistent where $V$ contains $2(n-1)+m+2 = 2n + m$ 
unlabelled inequations. $\Box$
\begin{theorem}[Complexity of $\mathbf{GA_l}$] \label{thmcomgal}
$\mathbf{GA_l} \in$ co-NP
\end{theorem}
\textbf{Proof.} Let $n$ be the size of the input formula $A$. To refute $A$ we choose a branch $B$ of the proof tree non-deterministically. $B$ is of $O(n)$ length (since each logical rule strictly reduces the number of connectives in the sequent) and ends with a labelled sequent $\Gamma \vdash \Delta$ with $O(n)$ atomic labels (since $(\limp,l)$ introduces exactly one new atomic label). So applying Lemma \ref{lemredseq} to the labelled and, by Proposition \ref{proplabreg} label-regular, inequation $\Gamma > \Delta$ we obtain a set of unlabelled inequations $U$ of $O(n)$ size in $O(n)$ time such that checking the $(success)$ rule for $\Gamma \vdash \Delta$ is equivalent to checking the consistency of $U$ over $\mathbb{Q}$ ie a linear programming problem. Since linear programming can be solved in polynomial time we are done. $\Box$
\subsection{A Labelled Single Sequent Calculus for \L}
Labelled sequents for \textbf{\L} are interpreted using the characteristic model $[-1,0]_\textrm{\L}$.
\begin{definition}[Interpretation of Labelled Sequents for \L]
$\models^*_\textrm{\L} \Gamma \vdash \Delta$ iff for all $v:For \rightarrow [-1,0]_\textrm{\L}$ there exists a labelling function $f$ such that $\Sigma_{A \in \Gamma} v(f(A)) \le \Sigma_{B \in \Delta} v(f(B))$.
\end{definition}
We also extend the translation * to labelled sequents.
\begin{definition}[Translation * for Labelled Sequents]
Given a labelled multiset $\Gamma$, $\Gamma^* = \{x:A^* | x:A \in \Gamma\}$. Given a labelled sequent $S = \Gamma \vdash \Delta$, $S^* = \Gamma^* \vdash \Delta^*$.
\end{definition}
\begin{theorem} \label{thmextlabtrans}
$\models^*_\textrm{\L} S$ iff $\models_A S^*$.
\end{theorem}
\textbf{Proof.} The proof of Theorem \ref{thmtrans} is easily extended from formulae to labelled sequents. $\Box$\\[.1in]
To give a suitable $(success)$ rule for \textbf{\L} we must first define a relation $\subseteq^*$ that takes account of internal weakening and $\bot$.
\begin{definition}[$\mathbf{\subseteq^*}$]
$\Delta \subseteq^* \Gamma$ iff (1) $\Delta \subseteq \Gamma$, or (2) $\Delta' \subseteq^* \Gamma'$ where $\Gamma = \Gamma' \cup \{\bot\}$ and $\Delta = \Delta' \cup \{q\}$ for some atom $q$.
\end{definition}
We now present the following labelled calculus for \textbf{\L}.
\begin{definition}[$\mathbf{G\textbf{\L}_l}$]
$\mathbf{G\textbf{\L}_l}$ has the following rules:
\[
\newcommand*{\lplusleft}{\begin{array}[t]{c}
\Gamma, x:A, x:B \vdash \Delta\\
\hline
\Gamma, x:A+B \vdash \Delta
\end{array}}
\newcommand*{\lplusright}{\begin{array}[t]{c}
\Gamma \vdash \Delta, x:A,x:B\\
\hline
\Gamma \vdash \Delta, x:A+B
\end{array}}
\newcommand*{\ltoleft}{\begin{array}[t]{c}
\Gamma,x:B \vdash \Delta,x:A\\
\hline
\Gamma,x:A \to B \vdash \Delta
\end{array}}
\newcommand*{\ltoright}{\begin{array}[t]{c}
\Gamma,x:A \vdash \Delta,x:B\\
\hline
\Gamma \vdash \Delta, x:A \to B
\end{array}}
\newcommand*{\llimpleft}{\begin{array}[t]{c}
\Gamma,xy:B \vdash \Delta,xy:A\\
\hline
\Gamma,x:A \limp B \vdash \Delta
\end{array}}
\newcommand*{\llimpright}{\begin{array}[t]{c}
\Gamma,x:A \vdash \Delta,x:B \ \ \Gamma \vdash \Delta\\
\hline
\Gamma \vdash \Delta, x:A \limp B
\end{array}}
\newcommand*{\lwk}{\begin{array}[t]{c}
\Gamma \vdash \Delta\\
\hline
\Gamma, x:A \limp B \vdash \Delta
\end{array}}
\newcommand*{\lsucc}{\begin{array}[t]{l}
\Gamma \vdash \Delta\\
\textnormal{where $\Gamma$ and $\Delta$ are atomic}\\
\textnormal{and there exist labelling}\\
\textnormal{functions $f_1, \ldots f_n$ such that}\\
\textnormal{$\cup_{i=1}^n f_i(\Delta) \subseteq^* \cup_{i=1}^n f_i(\Gamma)$}\\
\end{array}}
\small{
\begin{array}[t]{lclc}
(\limp,l) & \llimpleft & (\limp,r) & \llimpright\\
& \textnormal{$y$ a new atomic label} & &
\end{array}}
\]
\[
\small{
\begin{array}[t]{lcl}
(success) & \Gamma \vdash \Delta & \textnormal{where $\Gamma$ and $\Delta$ are atomic and there exist labelling}\\
& & \textnormal{functions $f_1, \ldots f_n$ such that $\cup_{i=1}^n f_i(\Gamma) \subseteq^* \cup_{i=1}^n f_i(\Delta)$}
\end{array}}
\]
\end{definition}
As in section 4, the soundness and completeness of \textbf{G\L}$_\mathbf{l}$ are proved proof-theoretically using the translation from \textbf{\L} to \textbf{A}.
\begin{proposition} \label{propsiffsstar}
$S$ succeeds in \textbf{G\L}$_\mathbf{l}$ iff $S^*$ succeeds in \textbf{GA}$_\mathbf{l}$.
\end{proposition}
\textbf{Proof.} Similar to (and in fact easier than) the proof of Proposition \ref{propgiffgstar}. $\Box$
\begin{theorem}[Soundness and Completeness of \textbf{G\L}$\mathbf{_l}$]
$S$ succeeds in \textbf{G\L}$\mathbf{_l}$ iff $\models^*_\textrm{\L} S$.
\end{theorem}
\textbf{Proof.} By Proposition \ref{propsiffsstar}, Theorems \ref{thmcomplab} and \ref{thmsoundlga}, and Theorem \ref{thmextlabtrans} respectively we have that $S$ succeeds in \textbf{G\L}$\mathbf{_l}$ iff $S^*$ succeeds in \textbf{GA}$\mathbf{_l}$ iff $\models_A S^*$ iff $\models^*_\textrm{\L} S$. $\Box$
\begin{theorem}[Complexity of \textbf{G\L}$\mathbf{_l}$]
\textbf{G\L}$\mathbf{_l}$ is co-NP-complete.
\end{theorem}
\textbf{Proof.} Similar to the proof of Theorem \ref{thmcomgal}. $\Box$
\section{Unlabelled Single Sequent Calculi}
In this section we obtain  \emph{unlabelled} single sequent calculi for 
\textbf{A} and \textbf{\L}. Instead of maintaining several components in a hypersequent or introducing labels we add rules permitting contraction from a sequent $\Gamma,\Gamma \vdash \Delta,\Delta$ to $\Gamma \vdash \Delta$, derivable in \textbf{GA} by applying $(EC)$ and $(S)$. We also alter the $(\limp,l)$ rules using the identity $A + (A \limp B) = B + (B \limp A)$, thereby obtaining sound and complete single sequent calculi for \textbf{A} and \textbf{\L}.
\subsection{An Unlabelled Single Sequent Calculus for A}
We present the following unlabelled single sequent calculus for \textbf{A}.
\begin{definition}[$\mathbf{GA_s}$]
$\mathbf{GA_s}$ has the following rules.\\[.1in]
\small{
Axioms
\[
\begin{array}[t]{lcclc}
(ID) & A \vdash A & \ \ \ \ \ \ & (\Lambda) \vdash
\end{array}
\]
Structural Rules
\[
\newcommand*{\tempd}{\begin{array}[t]{c}
\overbrace{\Gamma, \ldots, \Gamma}^n \vdash \overbrace{\Delta, \ldots, 
\Delta}^n\\
\hline
\Gamma \vdash \Delta
\end{array}}
\newcommand*{\tempk}{\begin{array}[t]{c}
\Gamma \vdash \Delta\\
\hline
\Gamma, A \limp B \vdash \Delta
\end{array}}
\newcommand*{\extmingle}{\begin{array}[t]{c}
\Gamma_1 \vdash \Delta_1 \ \Gamma_2 \vdash \Delta_2\\
\hline
\Gamma_1,\Gamma_2 \vdash \Delta_1,\Delta_2
\end{array}}
\begin{array}[t]{lllll}
(W) & \tempk & & (M) & \extmingle\\
(C) & \tempd & n>0 & &
\end{array}
\]
Logical Rules
\[
\newcommand*{\trightrule}{\begin{array}[t]{c}
\Gamma \vdash \Delta\\
\hline
\Gamma \vdash t, \Delta
\end{array}}
\newcommand*{\tleftrule}{\begin{array}[t]{c}
\Gamma \vdash \Delta\\
\hline
\Gamma,t \vdash \Delta
\end{array}}
\newcommand*{\arrowrightrule}{\begin{array}[t]{c}
\Gamma,A \vdash B,\Delta\\
\hline
\Gamma \vdash A \to B, \Delta
\end{array}}
\newcommand*{\arrowleftrule}{\begin{array}[t]{c}
\Gamma,B \vdash A,\Delta\\
\hline
\Gamma, A \to B \vdash \Delta
\end{array}}
\newcommand*{\plusrightrule}{\begin{array}[t]{c}
\Gamma \vdash A,B,\Delta\\
\hline
\Gamma \vdash A+B, \Delta
\end{array}}
\newcommand*{\plusleftrule}{\begin{array}[t]{c}
\Gamma,A,B \vdash \Delta\\
\hline
\Gamma, A+ B \vdash \Delta
\end{array}}
\newcommand*{\lognotleftrule}{\begin{array}[t]{c}
\Gamma \vdash A, \Delta\\
\hline
\Gamma, \lnot A \vdash \Delta
\end{array}}
\newcommand*{\lognotrightrule}{\begin{array}[t]{c}
\Gamma, A \vdash \Delta\\
\hline
\Gamma \vdash \lnot A, \Delta
\end{array}}
\newcommand*{\tempa}{\begin{array}[t]{c}
\Gamma \vdash \Delta \ \ \Gamma, A \vdash \Delta, B\\
\hline
\Gamma \vdash \Delta, A \limp B
\end{array}}
\newcommand*{\tempb}{\begin{array}[t]{c}
\Gamma, B, B \limp A \vdash \Delta, A\\
\hline
\Gamma, A \limp B \vdash \Delta
\end{array}}
\begin{array}[t]{lclc}
(t,l) & \tleftrule & (t,r) & \trightrule\\
& & &\\
(\lnot,l) & \lognotleftrule & (\lnot,r) & \lognotrightrule\\
& & &\\
(\to,l) & \arrowleftrule & (\to,r) & \arrowrightrule\\
& & &\\
(+,l) & \plusleftrule & (+,r) & \plusrightrule\\
& & &\\
(\limp,l) & \tempb & (\limp,r) & \tempa
\end{array}
\]
}
\end{definition}
Derived rules for $\land$ and $\lor$ are:
\[
\small{
\newcommand*{\rulex}{\begin{array}[t]{c}
\Gamma, A, A \limp B \vdash \Delta\\
\hline
\Gamma, A \land B \vdash \Delta\\
\end{array}}
\newcommand*{\ruley}{\begin{array}[t]{c}
\Gamma \vdash \Delta, A \ \ \Gamma \vdash \Delta, B\\
\hline
\Gamma \vdash \Delta, A \land B\\
\end{array}}
\newcommand*{\rula}{\begin{array}[t]{c}
\Gamma, A \vdash \Delta \ \ \Gamma, B \vdash \Delta\\
\hline
\Gamma, A \lor B \vdash \Delta\\
\end{array}}
\newcommand*{\rulb}{\begin{array}[t]{c}
\Gamma, B \limp A \vdash \Delta, A\\
\hline
\Gamma \vdash \Delta, A \lor B\\
\end{array}}
\begin{array}[t]{lclc}
(\land,l) & \rulex & (\land,r) & \ruley\\
& & &\\
(\lor,l) & \rula & (\lor,r) & \rulb\\
\end{array}}\]
To understand the differences between this calculus and \textbf{GA} compare the following proof of prelinearity with Example \ref{exama}.
\begin{example} 
A proof of $(A \to B) \lor (B \to A)$ in \textbf{GA}$\mathbf{_s}$:\\
\[\infer{\vdash (A \to B) \lor (B \to A)}{
	\infer{(B \to A) \limp (A \to B) \vdash A \to B}{
		\infer{(B \to A) \limp (A \to B), A \vdash B}{
			\infer{(B \to A) \limp (A \to B),(B \to A) \limp (A \to B), A, A \vdash B, B}{
				\infer{(B \to A) \limp (A \to B), A, A \vdash B, B}{
	\infer{(A \to B) \limp (B \to A),A \to B,A,A \vdash B,B,B \to A}{
		\infer{A \to B,A,A \vdash B,B,B \to A}{
			\infer{B,A,A \vdash B,B,B \to A,A}{
				\infer{B,B,A,A \vdash B,B,A,A}{
					\infer{B,B \vdash B,B}{
						B \vdash B & B \vdash B} &
					\infer{A,A \vdash A,A}{
						A \vdash A & A \vdash A}}}}}}}}}}
\]
\end{example}
Soundness is proved in the usual way.
\begin{theorem}[Soundness of $\mathbf{GA_s}$] \label{thmsounduga}
If $S$ succeeds in $\mathbf{GA_s}$ then $\models_A S$.
\end{theorem}
\textbf{Proof.} We follow the standard inductive proof and just check soundness in $\mathbb{Q}$ for the rules different to those of $\mathbf{GA}$.
\begin{itemize}
\item $(W)$. If $v(\Gamma) \le v(\Delta)$ then since $v(A \limp B) \le 0$ we have $v(\Gamma) + v(A \limp B) \le v(\Delta)$ for all valuations $v$.
\item $(C)$. If $\underbrace{v(\Gamma)+ \ldots + v(\Gamma)}_n \le 
\underbrace{v(\Delta) + \ldots + v(\Delta)}_n$ then dividing by $n$ gives 
$v(\Gamma) \le v(\Delta)$ for all valuations $v$.
\item $(\limp,l)$. For any valuation $v$, if $v(\Gamma) + v(B) + v(B \limp A) \le v(\Delta) + v(A)$ then since 
$v(B) + v(B \limp A) - v(A) = 
v(A \limp B)$ we have $v(\Gamma) + v(A \limp B) \le v(\Delta)$. $\Box$
\end{itemize}
We now turn our attention to proving the completeness of $\mathbf{GA_s}$. Our 
strategy is to show that any sequent provable in the \emph{labelled} calculus 
$\mathbf{GA_l}$ is provable in $\mathbf{GA_s}$. To this end we 
introduce an intermediate calculus $\mathbf{GA_i}$ which performs proofs in 
$\mathbf{GA_l}$ but which also maintains a record 
of the formulae added by the $(\limp,l)$ rule for $\mathbf{GA_s}$ in a \emph{store}; this allows us 
to prove inductively that any 
of the sequents represented by a labelled sequent $\Gamma\|\Pi \vdash \Delta$ (where $\Pi$ is the store) can 
be reached in $\mathbf{GA_s}$ 
from the unlabelled sequent $\Gamma^{ul},\Pi \vdash \Delta^{ul}$.
%
\begin{definition}[$\mathbf{\limp\!}$\textbf{-formula}, $\mathbf{atoms(\Gamma)}$, 
$\mathbf{nonatoms(\Gamma)}$] A 
$\limp$-formula is a formula of the form $A \limp B$, $atoms(\Gamma) = \{q \in 
\Gamma \ | \ q$ 
atomic$\}$, $nonatoms(\Gamma) = \{A 
\in \Gamma \ | \ A$ not atomic$\}$.
\end{definition}
\begin{definition}[$\mathbf{GA_i}$]
The rules for $\mathbf{GA_i}$ are exactly the same as for $\mathbf{GA_l}$ except 
that each premise and 
conclusion $\Gamma \vdash \Delta$ is replaced by $\Gamma\|\Pi \vdash \Delta$ and 
$(\limp,l)$ becomes:
\small{
\[
\newcommand*{\newrule}
{
\begin{array}[t]{c}
\Gamma, xy:B\|\Pi, B \limp A \vdash \Delta,xy:A\\
\hline
\Gamma, x:A \limp B\|\Pi \vdash \Delta
\end{array}
}
\newcommand*{\conds}
{
\begin{array}[t]{l}
(1) \ y \textnormal{ is a new atomic label.}\\
(2) \ \Delta \textnormal{ contains only atoms.}\\
(3) \ \Gamma \textnormal{ contains only atoms and} \limp\textnormal{-formulae.}
\end{array}
}
\begin{array}[t]{lll}
\newrule & \textnormal{where:} & \conds
\end{array}
\]
}
\end{definition}
We emphasize that formulae in the store are not used in $\mathbf{GA_i}$.
\begin{proposition} \label{propliffi}
$\Gamma \vdash \Delta$ succeeds in $\mathbf{GA_l}$ iff $\Gamma \|\Pi \vdash \Delta$ succeeds in $\mathbf{GA_i}$.
\end{proposition}
\textbf{Proof.}
We simply observe that the unlabelled formulae in the store are not processed by the rules of the calculus and 
that the restrictions for $(\limp,l)$ can easily be shown not to affect 
completeness. $\Box$
\begin{definition}[Branch]
A branch $[S_1,\ldots,S_n]$ in a calculus \textbf{G} is a sequence of sequents (or hypersequents) where for $i=1 \ldots n-1$, $S_i$ is the conclusion and $S_{i+1}$ a premise of a rule of \textbf{G}.
\end{definition}
\begin{proposition}\label{proprev}
Given a branch $[\Gamma, \Delta \vdash \Gamma, \Delta \ldots \Gamma,\Sigma^+ \vdash \Sigma^-,\Delta]$ in $\mathbf{GA_s}$ containing only applications of the logical rules excepting $(\limp,l)$, $\Gamma,\Sigma^+ \vdash \Sigma^-,\Delta$ succeeds in $\mathbf{GA_s}$.
\end{proposition}
\textbf{Proof.} By induction on the multiset complexity of $\Gamma \cup \Delta$. $\Box$\\[.1in]
We now prove the main technical proposition for our completeness proof. The idea is that at any point in a proof in $\mathbf{GA_i}$ where only sequents and $\limp$-formulae occur in the sequent, and given any labelling function, we are able to divide the current sequent $S=\Gamma \| \Pi \vdash \Delta$ into three parts; the first is the result of applying the labelling function to $S$, the second is a sequent that succeeds (removing labels) in $\mathbf{GA_s}$ and the third consists of $\limp$-formulae. 
\begin{proposition} \label{propreach}
Given a branch of a proof in $\mathbf{GA_i}$ $[\Gamma^l_1\|\emptyset 
\vdash \Delta^l_1 \ldots \Gamma\|\Pi \vdash \Delta]$ where $\Delta$ is atomic and 
$\Gamma$ contains only atoms 
and $\limp$-formulae, then for every labelling function $f$ there exist 
multisets $\Gamma_a$, $\Gamma_e$, 
$\Gamma_r$, $\Delta_a$ and $\Delta_e$ such that:\\[.1in]
(1) $\Gamma^{ul} = \Gamma_a \cup \Gamma_e \cup \Gamma_r$ and $\Delta^{ul} = 
\Delta_a \cup 
\Delta_e$\\
(2) $\Gamma_a = atoms(f(\Gamma))$ and $\Delta_a = f(\Delta)$\\
(3) $\Gamma_e, \Pi \vdash \Delta_e$ succeeds in $\mathbf{GA_s}$\\
(4) $\Gamma_e \subseteq \{A \ | \ x:A \in \Gamma$, $f(x)=0\}$\\
(5) $\Gamma_r$ contains only $\limp$-formulae
\end{proposition}
\textbf{Proof.}
By induction on the number of atomic labels, $n$, occurring in $\Gamma \| \Pi \vdash \Delta$. If $n=0$ then we take $\Gamma_a = 
atoms(\Gamma^{ul})$, $\Gamma_e = \emptyset$, $\Gamma_r=nonatoms(\Gamma^{ul})$, 
$\Delta_a = \Delta^{ul}$ and 
$\Delta_e = \emptyset$ and the conditions hold. For $n>0$ we consider the last 
introduction of a label via 
$(\limp,l)$:\\[.1in]
From $Q = \Gamma, x:A \limp B\|\Pi \vdash \Delta$ to:\\[.1in]
$Q'=\Gamma, xy:B\|\Pi,B \limp A \vdash \Delta, xy:A$\\[.1in]
followed by applications of the logical rules excepting $(\limp,l)$ giving:\\[.1in]
$Q'' = \Gamma, \Sigma^+\|\Pi, B \limp A \vdash \Delta, \Sigma^-$\\[.1in]
where $\Sigma^-$ contains only atoms, $\Sigma^+$ contains only atoms and 
$\limp$-formulae, and all formulae in $\Sigma^-$ and $\Sigma^+$ have label 
$xy$.\\[.1in]
Now for a given labelling function $f$ we apply the induction hypothesis to 
$Q$, obtaining multisets $\Gamma_a, \Gamma_e, \Gamma_r, \Delta_a$ and $\Delta_e$ 
as specified above. We seek 
$\Gamma'_a, \Gamma'_e, \Gamma'_r, \Delta'_a$ and $\Delta'_e$ suitable for $Q''$. 
There are two cases to 
consider:
\begin{itemize}
\item $f(x)=f(y)=1$.
We take $\Gamma'_a = \Gamma_a \cup atoms(\Sigma^+)^{ul}$, $\Gamma'_e = 
\Gamma_e$, $\Gamma'_r = \Gamma_r - \{A \limp B\} \cup nonatoms(\Sigma^+)^{ul}$, $\Delta'_a = \Delta_a 
\cup (\Sigma^-)^{ul}$ and 
$\Delta'_e = \Delta_e$.  We have: (1) $(\Gamma \cup \Sigma^+)^{ul} = \Gamma'_a 
\cup \Gamma'_e \cup \Gamma'_r$ 
and $(\Delta \cup \Sigma^-)^{ul} = \Delta'_a \cup \Delta'_e$, (2) $\Gamma'_a = 
atoms(f(\Gamma \cup \Sigma^+))$ 
and $\Delta'_a = f(\Delta \cup \Sigma^-)$ since $f(xy)=1$, (3) $\Gamma'_e, \Pi 
\vdash \Delta'_e$ succeeds in 
$\mathbf{GA_s}$  and hence so does $\Gamma'_e,\Pi,B \limp A \vdash \Delta'_e$ by 
$(W)$, (4) $\Gamma'_e 
\subseteq \{A \ | \ x:A \in \Gamma \cup \Sigma^+$, $f(x)=0\} = \{A \ | \ x:A \in 
\Gamma, f(x)=0\}$ and (5) 
$\Gamma'_r$ contains only $\limp$-formulae.
\item $f(x)=0$ or $f(y)=0$.
We take $\Gamma'_a = \Gamma_a$, $\Delta'_a = \Delta_a$ and $\Delta'_e = 
\Delta_e \cup (\Sigma^-)^{ul}$ and then for $A \limp B \in \Gamma_e$, we take 
$\Gamma'_e = \Gamma_e - \{A \limp B\} \cup (\Sigma^+)^{ul}$ and 
$\Gamma'_r = \Gamma_r$, for $A \limp B 
\in \Gamma_r$, we take $\Gamma'_e = \Gamma_e 
\cup (\Sigma^+)^{ul}$ and $\Gamma'_r = \Gamma_r - \{A \limp B\}$. 
We have: (1) $(\Gamma \cup \Sigma^+)^{ul} = \Gamma'_a \cup \Gamma'_e \cup 
\Gamma'_r$ and $(\Delta \cup 
\Sigma^-)^{ul} = \Delta'_a \cup \Delta'_e$ by definition; (2) $\Gamma'_a = 
atoms(f(\Gamma \cup \Sigma^+))$ and 
$\Delta'_a = f(\Delta \cup \Sigma^-)$ since $f(xy)=0$; (3) we have that 
$\Gamma_e, \Pi \vdash \Delta_e$ succeeds in $\mathbf{GA_s}$ so (for $A \limp B 
\in \Gamma_e$) by Proposition \ref{proprev} 
$\Gamma_e - \{A \limp B\}, (\Sigma^+)^{ul}, \Pi, B \limp A \vdash 
\Delta_e, (\Sigma^-)^{ul}$ succeeds in $\mathbf{GA_s}$, and (for $A \limp B \not 
\in \Gamma_r$), since $B \limp A, (\Sigma^+)^{ul} \vdash (\Sigma^-)^{ul}$ 
succeeds in $\mathbf{GA_s}$ using Proposition \ref{proprev} so also by $(M)$ 
does $\Gamma_e, B \limp A, (\Sigma^+)^{ul}, \Pi \vdash \Delta_e, 
(\Sigma^-)^{ul}$; (4) if 
$u:C \in \Gamma_e \cup 
(\Sigma^-)^{ul}$ then $f(u)=0$ so $\Gamma'_e \subseteq \{C | u:C \in \Gamma \cup 
\Sigma^+, f(u)=0\}$; (5) 
$\Gamma'_r$ contains only $\limp$-formulae. $\Box$
\end{itemize}
We now use this proposition to prove the completeness of $\mathbf{GA_s}$.
\begin{theorem}[Completeness of $\mathbf{GA_s}$] \label{thmcompulga}
If $\models_A S$ then $S$ succeeds in $\mathbf{GA_s}$.
\end{theorem}
\textbf{Proof.} If $\models_A \Gamma_1 \vdash \Delta_1$ then by Theorem \ref{thmcomplab} we have that $\Gamma^l_1 \vdash \Delta^l_1$ succeeds in $\mathbf{GA_l}$ and hence by Proposition \ref{propliffi} $\Gamma^l_1 \| \emptyset \vdash \Delta^l_1$ also succeeds in $\mathbf{GA_i}$. Given a proof of $\Gamma^l_1 \| \emptyset \vdash \Delta^l_1$ in $\mathbf{GA_i}$ we apply the corresponding rules of $\mathbf{GA_s}$ until the point on each branch when $(success)$ is applied to a sequent $\Gamma\|\Pi \vdash \Delta$ ie where there are labelling functions $f_1, \ldots, f_n$ such that $\cup_{i=1}^n f_i(\Gamma) = \cup_{j=1}^n f_j(\Delta)$. Now by Proposition \ref{propreach}, for $i=1 \ldots n$ there 
exist multisets $\Gamma^i_a, 
\Gamma^i_e, \Gamma^i_r, \Delta^i_a, \Delta^i_e$ such that:\\[.1in]
(1) $\Gamma^{ul} = \Gamma^i_a \cup \Gamma^i_e \cup \Gamma^i_r$
and $\Delta^{ul} = \Delta^i_a \cup \Delta^i_e$\\
(2) $\Gamma^i_a = atoms(f_i(\Gamma)) = f_i(\Gamma)$ and $\Delta^i_a = 
f_i(\Delta)$\\
(3) $\Gamma_e^i, \Pi \vdash \Delta_e^i$ succeeds in $\mathbf{GA_s}$\\
(4) $\Gamma^i_e \subseteq \{A \ | \ x:A \in \Gamma$, $f_i(x)=0\}$\\
(5) $\Gamma^i_r$ contains only $\limp$-formulae\\[.1in]
We show that the corresponding sequent $\Gamma^{ul}, \Pi \vdash 
\Delta^{ul}$ succeeds in 
$\mathbf{GA_s}$. First we use $(C)$ to step to:\\[.1in]
$\underbrace{\Gamma^{ul}, \ldots, \Gamma^{ul}}_n, \underbrace{\Pi, \ldots, 
\Pi}_n \vdash 
\underbrace{\Delta^{ul}, \ldots, \Delta^{ul}}_n$\\[.1in]
We then apply (W) repeatedly to formulae in $\Gamma^i_r$ for $i=1 \ldots n$ 
obtaining:\\[.1in]
$\Gamma^1_a \cup \Gamma^1_e, \ldots, \Gamma^n_a \cup 
\Gamma^n_e, \underbrace{\Pi, \ldots, \Pi}_n 
\vdash \Delta^1_a \cup \Delta^1_e, \ldots, \Delta^n_a \cup 
\Delta^n_e$\\[.1in]
Since $\cup_{i=1}^n f_i(\Delta) = \cup_{j=1}^n f_j(\Gamma)$ we have 
that $\Gamma^1_a, \ldots, \Gamma^n_a \vdash \Delta^1_a, \ldots, \Delta^n_a$ 
succeeds in $\mathbf{GA_s}$ using 
$(M)$ and $(ID)$. Also $\Gamma^i_e,\Pi \vdash \Delta^i_e$ succeeds in $\mathbf{GA_s}$ for 
$i=1 \ldots n$. So applying $(M)$ 
repeatedly, we get that the whole sequent succeeds as required. $\Box$
\subsection{An Unlabelled Single Sequent Calculus for \L}
We present the following unlabelled single sequent calculus for \textbf{\L}.
\begin{definition}[$\mathbf{G\textbf{\L}_s}$]
$\mathbf{G\textbf{\L} _s}$ has the following rules.\\[.1in]
\small{
Axioms
\[
\begin{array}[t]{lcclcclc}
(ID) & A \vdash A & \ \ \ \ \ \ & (\Lambda) & \vdash & \ \ \ \ \ \ & (\bot) & \bot \vdash A
\end{array}
\]
Structural Rules
\[
\newcommand*{\tempd}{\begin{array}[t]{c}
\overbrace{\Gamma, \ldots, \Gamma}^n \vdash \overbrace{\Delta, \ldots, 
\Delta}^n\\
\hline
\Gamma \vdash \Delta
\end{array}}
\newcommand*{\tempk}{\begin{array}[t]{c}
\Gamma \vdash \Delta\\
\hline
\Gamma, A \vdash \Delta
\end{array}}
\newcommand*{\extmingle}{\begin{array}[t]{c}
\Gamma_1 \vdash \Delta_1 \ \Gamma_2 \vdash \Delta_2\\
\hline
\Gamma_1,\Gamma_2 \vdash \Delta_1,\Delta_2
\end{array}}
\begin{array}[t]{lllll}
(W) & \tempk & & (M) & \extmingle\\
(C) & \tempd & n>0 & &
\end{array}
\]
Logical Rules
\[
\newcommand*{\tempa}{\begin{array}[t]{c}
\Gamma \vdash \Delta \ \ \Gamma, A \vdash \Delta, B\\
\hline
\Gamma \vdash \Delta, A \limp B
\end{array}}
\newcommand*{\tempb}{\begin{array}[t]{c}
\Gamma, B, B \limp A \vdash \Delta, A\\
\hline
\Gamma, A \limp B \vdash \Delta
\end{array}}
\begin{array}[t]{lclc}
(\limp,l) & \tempb & (\limp,r) & \tempa
\end{array}
\]
}
\end{definition}
We give the following proof of the characteristic axiom of \textbf{\L} for comparison with the \textbf{G\L} proof in Example \ref{examb}.
\begin{example}
A proof of $((A \limp B) \limp B) \limp ((B \limp A) \limp A)$ in \textbf{G\L}$\mathbf{_l}$:
\small{
\[
\infer{\vdash ((A \limp B) \limp B) \limp ((B \limp A) \limp A)}{
	\vdash &
	\infer{(A \limp B) \limp B \vdash (B \limp A) \limp A}{
		\infer{(A \limp B) \limp B \vdash}{
			\vdash} &
		\infer{(A \limp B) \limp B, B \limp A \vdash A}{
			\infer{(A \limp B) \limp B, A \limp B, A \vdash A, B}{
				A \vdash A &
				\infer{(A \limp B) \limp B, A \limp B \vdash B}{
					\infer{B \limp (A \limp B), B, A \limp B \vdash B, 
A \limp B}{
					\infer{B, A \limp B \vdash B, A \limp B}{
						B \vdash B &
						A \limp B \vdash A \limp B}}}}}}}
\]}

\end{example}
As usual the soundness and completeness of \textbf{G\L}$\mathbf{_s}$ is proved proof-theoretically using the translation from \textbf{\L} to \textbf{A}.
\begin{proposition} \label{propsuliffsulstar}
$S$ succeeds in \textbf{G\L}$\mathbf{_s}$ iff $S^*$ succeeds in $\mathbf{GA_s}$.
\end{proposition}
\textbf{Proof.} For the left-to-right direction it is straightforward to show that the translated rules of \textbf{G\L}$\mathbf{_s}$ are either derivable or admissible in $\mathbf{GA_s}$. For the right-to-left direction we just sketch a proof. First we replace the axioms and the rules $(W)$ and $(M)$ of \textbf{G\L}$\mathbf{_s}$ with $(AX)$ $\Gamma,t \ldots,t \vdash \Delta,t \ldots,t$ where $\Delta \subseteq^* \Gamma$. It is easy to show that the two versions prove the same theorems (replacing $t$ with $q \limp q$ where appropriate). Now since the rules and axioms of $\mathbf{GA_s}$ applying to $\limp$ and propositional variables are a subset of those for \textbf{G\L}$\mathbf{_s}$ we have that if $S^*$ succeeds in $\mathbf{GA_s}$ then $S^*$ succeeds in \textbf{G\L}$\mathbf{_s}$. Hence we can then prove inductively that if $S^*$ succeeds in \textbf{G\L}$\mathbf{_s}$ then $S$ succeeds in \textbf{G\L}$\mathbf{_s}$ and we are done. $\Box$
%
%
\begin{theorem}[Soundness and Completeness of \textbf{G\L}$\mathbf{_s}$]
$S$ succeeds in \textbf{G\L}$\mathbf{_s}$ iff $\models^*_\textrm{\L} S$.
\end{theorem}
\textbf{Proof.} By Proposition \ref{propsuliffsulstar}, Theorems \ref{thmcompulga} and \ref{thmsounduga}, and Theorem \ref{thmexttrans} respectively we have that $S$ succeeds in \textbf{G\L}$\mathbf{_s}$ iff $S^*$ succeeds in $\mathbf{GA_s}$ iff $\models_A S^*$ iff $\models^*_\textrm{\L} S$. $\Box$
\section{Discussion}
In this paper we have presented the first Gentzen-style analytic and internal sequent and hypersequent calculi for abelian logic \textbf{A} and \L ukasiewicz infinite-valued logic \textbf{\L}. We have also developed terminating versions of the hypersequent calculi and co-NP labelled sequent calculi matching the complexity bounds for both logics. The key step for achieving these results for \textbf{\L} has been a translation of \textbf{\L} into \textbf{A}. 

For abelian logic \textbf{A} our work, providing the first comprehensive proof systems for this logic, extends that of Paoli \cite{pao:log} who gave a calculus for just the intensional part of \textbf{A}. For \L ukasiewicz logic \textbf{\L}, on the other hand, several proof systems have been presented in the literature. All of these however either fail to be analytic, that is they have cut as a uneliminable rule eg \cite{Prijatelj96,Ciabattoni:1997:TCB}, or they rely on external non-logical calculations such as solving mathematical programming problems \cite{hahnle:admvl,mun:res,oli:tab}. One exception is the proof calculus of Aguzzoli and Ciabattoni \cite{AguzzoliCiabattoni99} which exploits the fact that any formula valid in \textbf{\L} is also valid in \textbf{\L}$\mathbf{_n}$ where $n$ is a function of the number of occurrences of variables in the formula. This calculus uses multiple sequents that seem to perform a similar role proof-theoretically to hypersequents but with a very different semantic interpretation. Although this approach provides a valuable perspective on the relationship between \textbf{\L} and the finite-valued logics we take the view that \textbf{\L} both can and should be viewed as \emph{independent} of the finite-valued logics. This is supported by the fact that the calculi given in this paper, in particular \textbf{G\L} and \textbf{G\L}$\mathbf{_s}$, are more direct and natural than those for \textbf{\L}$\mathbf{_n}$. Moreover proof systems for the finite-valued logics may be obtained from our systems for \textbf{\L} by adding appropriate extra axioms or rules.

A more implementation-oriented approach to \L ukasiewicz logic is introduced by H\"ahnle in \cite{hahnle:admvl} where a \emph{labelled tableaux} reduction of \textbf{\L} to \emph{mixed integer programming} is presented. Although similar in output (ie mathematical programming problems) to our calculus \textbf{G\L}$\mathbf{_t}$ and of the same complexity (co-NP) the method is significantly different to ours in that constraints (ie equations) are generated dynamically as a proof progresses rather than maintaining a single labelled equation for each branch. For this reason our calculus seems to be more \emph{logical} and easier to understand at various stages of a proof. A further tableaux calculus has also been presented by Olivetti in \cite{oli:tab} based on the Kripke semantics of \textbf{\L} that again performs a reduction to mathematical programming problems and is co-NP.

Finally we note that our approach of embedding fuzzy logics in comparative logics could prove successful in deriving sequent and hypersequent calculi for the other fuzzy logics. In particular we would like to find proof systems for H\'ajek's basic logic \textbf{BL} and Product logic \textbf{P}. 
\bibliographystyle{plain}
\bibliography{bibliography}
\end{document}